\documentclass[%
aip,
amsmath,amssymb,
preprint,%
floatfix
]{revtex4-1}

\usepackage{graphicx}
\usepackage{dcolumn}
\usepackage{bm}
\usepackage[utf8]{inputenc}
\usepackage[T1]{fontenc}
\usepackage{mathptmx}
\usepackage{etoolbox}

\makeatletter
\def\@email#1#2{%
 \endgroup
 \patchcmd{\titleblock@produce}
  {\frontmatter@RRAPformat}
  {\frontmatter@RRAPformat{\produce@RRAP{*#1\href{mailto:#2}{#2}}}\frontmatter@RRAPformat}
  {}{}
}%
\makeatother

\begin{document}

\noindent{\small{\texttt{The following article has been submitted to Structural Dynamics. After it is published, it will be found at https://publishing.aip.org/resources/librarians/products/journals/}}}

\vspace*{4\baselineskip}

\title{GARFIELD, a toolkit for interpreting ultrafast electron diffraction data of imperfect  quasi-single crystals}

\author{Alexander Marx*}
\email{alexander.marx@mpsd.mpg.de}
\author{Sascha W. Epp}
\affiliation{ 
Max Planck Institute for the Structure and Dynamics of Matter, CFEL, Luruper Chaussee 149, 22761 Hamburg, Germany}

\date{\today}

\begin{abstract}
The analysis of ultrafast electron diffraction (UED) data from low-symmetry single crystals of small molecules is often challenged by the difficulty of assigning unique Laue indices to the observed Bragg reflections. For a variety of technical and physical reasons, UED diffraction images are typically of lower quality when viewed from the perspective of structure determination by single-crystal X-ray or electron diffraction. Nevertheless, time series of UED images can provide valuable insight into structural dynamics, provided that an adequate interpretation of the diffraction patterns can be achieved. \textsc{Garfield} is a collection of tools with a graphical user interface designed to facilitate the interpretation of diffraction patterns and to index Bragg reflections in challenging cases where other indexing tools are ineffective. To this end, \textsc{Garfield} enables the user to interactively create, explore, and optimize sets of parameters that define the diffraction geometry and characteristic properties of the sample. 
\end{abstract}

\maketitle

\section{\label{sec:intro}Introduction}

Crystal structure determination by single-crystal X-ray or electron diffraction is usually based on the collection of a series of diffraction images, each containing a number of sharp and distinct reflections. Typically, the first step in diffraction data analysis is some form of "interpretation" of the diffraction patterns, which leads to an enormous reduction in the amount of data. The goal is to identify isolated reflection peaks and assign them to reciprocal lattice points (RLP) of a perfect crystal with Laue indices \textit{hkl}, and further to use the measured (partial) intensities to obtain estimates of the underlying full Bragg intensities.  These full intensities are -- in the simplest cases -- proportional to $|F_{hkl}|^2$, the squared moduli of the structure factors. The indexing task can be demanding, but powerful auto-indexing tools have been developed for various use cases and incorporated into commonly used crystallographic software packages.\cite{Kabsch1988a, Duisenberg1992, Kabsch1993, Powell1999, Lauridsen2001, Sauter2004, Kabsch2010, Powell2013, Gildea2014, Sauter2014, Schmidt2014, Ginn2016, Beyerlein2017, Gevorkov2019, Li2019, Gevorkov2020} They all amount to some form of spatial or geometric analysis of tabulated peak positions. For large unit cells, a single diffraction pattern can be sufficient to determine the orientation of the diffracting crystal and uniquely assign the reflection indices, especially with prior knowledge of the unit cell. This is of great benefit for serial femtosecond crystallography (SFX) and serial synchrotron crystallography (SSX) of macromolecular crystals where thousands of diffraction images, each taken from a different crystal, can be processed in a short time and with a high success rate.\cite{White2012, Hattne2014, Kabsch2014, Winter2018}

Due to their small wavelength and strong interaction with matter, the diffraction of electrons opens up new possibilities, but also presents new challenges. The contribution of multiple scattering to the diffraction signal can be significant, necessitating a dynamical approach rather than a kinematical one. The samples must be sufficiently thin, typically on the order of 100 nanometers or less, depending on the atomic composition. Moreover, the intrinsic small wavelength renders the indexing of individual diffraction images more difficult, as each image only encompasses information from a nearly planar section through reciprocal space. 
To address these challenges, a range of innovative electron diffraction protocols have been developed in modern electron microscopy. The most crucial aspect is the limitation of the probe volume to thin and flawless regions of the sample through the use of selected area electron diffraction (SAED) or nanobeam diffraction (NBD). In recent years, notable advancements have been made in the field of crystal structure determination by "three-dimensional electron diffraction" \cite{Gemmi2019}, both in material science and macromolecular crystallography. The advent of new methods and automated protocols, including microcrystal electron diffraction (MicroED)\cite{Shi2013}, precession electron diffraction (PED) \cite{Palatinus2015}, and parallel NBD combined with STEM imaging, has considerably alleviated the adverse influences of radiation damage as well as multiple scattering and facilitated indexing.   
The continuous expansion of the range of single-crystal diffraction data that can be indexed automatically has made serial electron crystallography with nanocrystalline materials \cite{Smeets2018} and protein nanocrystals \cite{Buecker2021} a feasible undertaking.

Ultrafast electron diffraction of bright femtosecond electron pulses (UED) \cite{Siwick2003, Ruan2004, Zewail2006, Sciaini2011, Miller2014, Ischenko2017, deCotret2018, Yang2020, Lee2024} for the analysis of structural dynamics poses unique requirements for the interpretation of the diffraction data. The goal of a UED experiment is to study structural changes after perturbation of the sample, e.g. by a short laser pulse at time $t_0$, by recording a series of diffraction images at times $t_i = t_0 + \Delta t_i$ and comparing them with images recorded without perturbation ($\Delta t_i < 0$). The data acquisition must be at a sufficient signal-to-noise level to allow accurate measurement of differences in diffraction intensities due to potentially small structural changes. UED data collected from thin, single-crystal samples typically consist of one or several time series of diffraction images, each series comprising a large number of nearly identical diffraction patterns. Thus, "indexing" one representative image per time series would be sufficient to index all members of the series. The challenges of interpreting such UED data often begin with the problem of assigning the correct Laue indices to the observed reflections since the observed data often do not look sufficiently similar to expectations derived from oversimplified approaches that are ultimately unsuitable for describing this type of diffraction data.

\begin{figure*}[htbp]
\centering
\includegraphics[width=\textwidth]{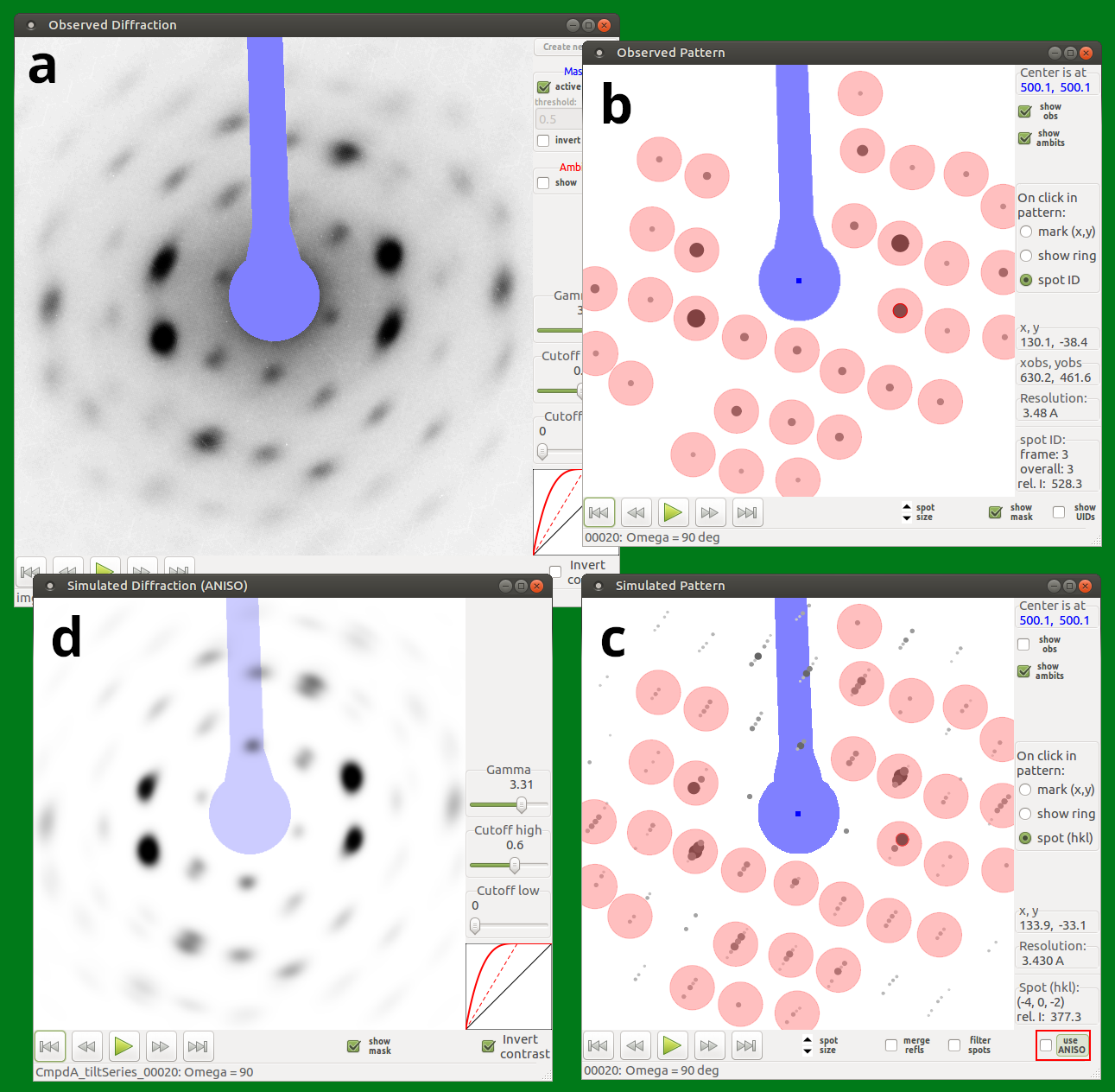}
\caption{Types of diffraction images and patterns available from \textsc{Garfield}'s Display menu.  \textbf{(a)} Diffraction image with mask (blue, beamstop). \textbf{(b)} Diffraction pattern (schematic representation of the \textit{reduced data}): observed diffraction spots are symbolized by dark discs with sizes and gray levels decreasing with intensity; here, overlaid with red discs showing the spot \textit{ambits} in use (see section \ref{sec:model.1} for explanations). Weak diffuse reflections correspond to small discs of low density, which can be counter-intuitive. \textbf{(c)} Simulated pattern analogous to (b), with observed spots replaced by calculated Bragg reflections. Reflections within the same ambit can be merged to a single spot for easier comparison with (b). Simulation can be done with either standard or ANISO modeling (see \ref{sec:model.4}). As in the example case, there is often no significant difference between the results of the two methods. \textbf{(d)}: Simulated diffraction image, calculated with the same parameters as used for (c). Simulation of diffraction images is always done with ANISO modeling. 
\big(Figure prepared with data from Ishikawa et al. (2015)\cite{Ishikawa2015}, HT phase of $\mathrm{Me_{4}P[Pt(dmit)_{2}]_{2}}$\big) 
}
\label{fig:CmpdA_combi}
\end{figure*}

The generation of temporally short and laterally confined high-fluence electron pulses inevitably leads to space charge effects that ultimately limit beam coherence.\cite{Chatelain2012} In order to mitigate these effects, the electron beam size at the sample in UED is usually quite large, typically on the order of 100~µm in diameter.  The combined requirement of a small sample thickness for high electron transmission and a large area to overlap with the beam dimensions makes UED samples susceptible to bending, cracking, corrugation, and other effects that degrade the sample quality and thus the quality of the diffraction data. Hence, most UED-ready crystalline samples are far from perfect single crystals, and are better described as an ensemble of crystalline domains or crystallites with similar but not identical orientation.  In addition, the preparation of UED samples, e.g. by thin-slicing crystals, often results in a high density of lattice defects, which reduces the average size of coherently diffracting domains and increases the effective size of reciprocal lattice points (RLPs) in all three spatial directions. For these technical and physical reasons, the Bragg reflections recorded in UED experiments are typically blurred and often overlap with other reflections, so that purely geometry-based indexing, which relies on precise peak positions, cannot be regularly applied. Typical UED data are of lower quality if judged from the viewpoint of well-established techniques of crystal structure analysis by X-ray or electron diffraction. An example from UED is shown in Figure \ref{fig:CmpdA_combi}a.
For the determination of static crystal structures, such images are usually excluded from further consideration at an early stage of the standard workflows, e.g. in the “crystal screening” phase of single crystal structure analysis or by fast and efficient filtering and restriction of indexing to “hits” in SFX and SSX. However, despite their shortcomings, time series of UED images can provide valuable information on structural dynamics, making the interpretation of such data highly desirable. Fortunately, the number of diffraction images to be indexed is usually small, namely one per time series. Therefore, a different approach is required to overcome the challenges of indexing typical UED data.

The \textsc{Garfield} software package, which is the subject of this report, has been developed with the objective of facilitating the step of indexing single-crystal UED data under challenging conditions. In essence, the software simulates electron diffraction within a model that is computationally efficient and fast while retaining the relevant features found in typical UED diffraction patterns. The software provides tools for determining model parameters (including orientation of the crystal lattice) that, according to reasonable criteria, best describe the observed diffraction data. It differs from available indexing tools in several aspects. The most distinct differences are: (1) the use of reflection intensities and not only peak positions in the analysis of diffraction data, and (2) the use of a graphical user interface allowing direct and interactive control of the evaluation process by breaking out of the common data-in-result-out black-box approach. The \textsc{Garfield} approach to interpreting and indexing diffraction data is an interactive process. The objective is to identify the "best" solution, which is approached step by step. This involves excluding unlikely solutions and optimizing promising solutions through non-linear least-squares (NLS) fitting of model parameters that describe diffraction geometry and sample properties. The two core tools in this process are \textsc{GridScan}, performing fast grid searches of possible orientations in three-dimensional rotation space, and \textsc{GeoFit}, a flexible tool for NLS fitting of individual parameters and whole parameter sets with presently up to 22 parameters for a single diffraction pattern. These include the (mean) orientation of the crystal lattice, the beam center, intensity and spatial scaling, sample thickness, and the orientation distribution of crystal domains.

The paper is structured as follows: Section \ref{sec:design} formulates the two guiding principles used to design the \textsc{Garfield} software package and describes how these principles are reflected in the chosen approach for indexing typical single-crystal UED data.
Section \ref{sec:work} summarizes the technical requirements for installing \textsc{Garfield} (A), provides some general instructions and recommendations for working with the graphical user interface (B, C), and introduces the core tools, \textsc{GridScan} (D) and \textsc{GeoFit} (E). 
Section \ref{sec:model}  describes the underlying model used to predict the position and intensity of diffraction spots for a given set of model parameters.
In section \ref{sec:conclusion} (Conclusion), the essential characteristics of \textsc{Garfield} are summarized, its benefits and limitations are discussed, and ways to improve its performance are outlined.

Although \textsc{Garfield} was developed with UED applications in mind and tested with data from low-symmetry molecular crystals, it can be useful in a broader range of applications where conventional methods for indexing fail. Additional examples of the application of \textsc{Garfield} can be found in the Supplementary Material.

\section{\label{sec:design}Design principles of \textsc{Garfield}}

Indexing single-crystal diffraction data is primarily a task of unraveling the geometric relationship between the probe beam direction, crystal lattice orientation, and detector geometry, with challenges posed by imperfect beam and crystal conditions coupled with detector insufficiency. Conventional approaches rely on the precise determination of the positions of the recorded Bragg peaks, whereas reflection intensities are mostly disregarded, except for the precise localization of reflections and the discrimination of proper reflections from artifacts by considering peak profiles (post-refinement is out of scope in the present context). If the the unit cell parameters are known and accurate positions of the Bragg peaks have been extracted from a diffraction image, the orientation of the crystal lattice can in many cases be calculated within a fraction of a second by using one of the readily available indexing tools. 

Since UED experiments typically lack precise information about the position of individual Bragg peaks, \textsc{Garfield}'s first design principle (P1) is to use as much of the available information as possible. Hence, the discriminating information in reflection intensities should not be ignored. However, the application of this principle has consequences that go well beyond the inclusion of intensity data, as discussed below. The second principle (P2) requires that the fuzziness of recorded diffraction data (due to imperfections of the sample, for instance) should be taken into account in the process of finding the best interpretation of the data. First, it can provide additional discriminating information, and second, it allows the (mean) orientation of the crystal lattice to be fitted with standard least-squares methods, which would not be possible in the case of data from a perfect crystal.

\textsc{Garfield}'s basic input must be provided as a list of detected diffraction coordinates $(x_{i_\mathrm{o}}^{\mathrm{obs}}, y_{i_\mathrm{o}}^{\mathrm{obs}})$ and partial intensities $I_{i_\mathrm{o}}^{\mathrm{obs}}$ (hereinafter called \textit{reduced data}) as well as a crystal structure file in the CIF format \cite{Bernstein2016}. Hence, the program can only be used if the crystal structure (of the unperturbed crystal) is already known. \textsc{Garfield} calculates predictions $(x_{i_\mathrm{o}}^{\mathrm{clc}}, y_{i_\mathrm{o}}^{\mathrm{clc}})$ and $I_{i_\mathrm{o}}^{\mathrm{clc}}$ within a simple model in kinematical approximation ($i_\mathrm{o} = 1, ...~N_\mathrm{obs}$, where $N_\mathrm{obs}$ is the number of Bragg spots \footnote{A \textit{Bragg spot} can be a single Bragg reflection or the superposition of several Bragg reflections.}, or more precisely, the number of rows in the list of reduced data). Besides physical and geometrical parameters (such as wavelength $\lambda$, detector distance, rotation angles $\Theta, \Phi, \Psi$ describing the orientation of the crystal lattice, etc.) the model includes parameters for quantifying several effects that affect the quality of recorded UED diffraction images: beam divergence, the finite size of the coherently diffracting domains, and -- most important for UED -- the domain orientation distribution. The latter will be referred to as \textit{mosaicity}, although this may be an extension of the usual meaning. Using the \textsc{GeoFit} tool, almost all parameters, including the orientation of the crystal lattice, can be optimized by simultaneous NLS fitting. This is possible, because the cost function $S$, the function to be minimized (the sum of squared residuals, SSR, essentially),
\begin{equation} \label{eq1}
S  = S^\mathrm{(I)} ~ + ~ w_\mathrm{P} ~ S^\mathrm{(P)}, ~~ (w_\mathrm{P} \ge 0) 
\end{equation}
which for a perfect crystal under ideal conditions is a discontinuous function of the orientation parameters, is smoothed and becomes a continuous function under imperfect conditions as a consequence of the fuzziness due to various blurring effects. In equation \ref{eq1}, $w_\mathrm{P}$ is a control parameter that weights the discrepancy between predicted and observed peak positions, $S^\mathrm{(P)}$, relative to the intensity-dependent part $S^\mathrm{(I)}$ of the cost function. Using NLS fitting of positions \emph{and} intensities in determining the orientation of the crystal lattice is one of the characteristic features that distinguish \textsc{Garfield} from other tools. 

NLS fitting of 20 or more parameters, individually or simultaneously, is challenging. Convergence is not guaranteed, and starting values not too far from the solution are required in order to avoid convergence towards a wrong local minimum. Unsupervised fitting is therefore not practical under such conditions. Hence, \textsc{Garfield}'s workflow is not designed to work like a black-box which automatically returns aresult when a set of input data is entered, but rather provides a workbench that allows the user to work with multiple parameter sets and interactively explore the parameter space until a satisfactory result is found. This is the second fundamental difference between \textsc{Garfield} and other indexing tools. In essence, \textsc{Garfield} displays and manages a table of parameter sets, that can be modified and extended by the user. In the following, this table will be referred to as the "main table of settings", or the \emph{main table} for short. \textsc{Garfield} provides various tools to create and modify settings, either by manual editing, by least-squares optimization of existing parameter sets (\textsc{GeoFit}), or by systematic screening of crystal lattice orientations (\textsc{GridScan}). The goal is to arrive at a setting that optimally accounts for the totality of the recorded diffraction information. 

The quality of different parameter sets can be evaluated by SSR values and crystallographic R-factors (after NLS optimization), and by visual comparison of \textsc{Garfield}'s predicted diffraction with the observed diffraction.
Although not strictly quantitative, the importance of visual comparison cannot be overstated, since it is often the prediction of fine details in the recorded diffraction, not captured by the input list of reduced data, that makes a particular parameter set most convincing. This can be viewed as a special kind of cross-validation and is consistent with the first design principle mentioned above (P1), which would be violated if only the reduced data were considered. 

\section{Working with \textsc{Garfield}} \label{sec:work}

\subsection{Installing \textsc{Garfield}} \label{sec:work.1}

\textsc{Garfield} consists of a collection of scripts written in the \textsf{R} programming language for statistical computing \cite{Rstats}. \textsf{R} is a free software available for common computer platforms. \textsc{Garfield} needs installation of \textsf{R} and additional packages available from the Comprehensive R Archive Network (CRAN). Although most of the code is platform-independent, \textsc{Garfield} currently runs only on a Linux system with the GTK2 and Cairo libraries installed. Detailed step-by-step installation instructions are included in the documentation/user manual provided in form of a TiddlyWiki \cite{TiddlyWiki}, the file \texttt{Garfield\_wiki.html}, which is part of the \textsc{Garfield} distribution.

\subsection{Setting up a new project} \label{sec:work.2}

A \textsc{Garfield} \textit{project} is defined by (1) a table of reflection positions and intensities extracted from either a single diffraction image to be indexed, or a tilt series of diffraction images, (2) the crystal structure (in form of a CIF file) which must be known \textit{a priori}, and (3), optionally, a diffraction image (one or serveral in the case of a tilt series) in TIFF, JPEG or PNG format for visual comparison and verification. Here and in the following, the simplest case of a project for indexing a single diffraction image will be assumed. The generalization to multi-image projects, e.g. for tilt-series obtained by rotating the sample about an axis perpendicular the incident beam, is self-explanatory. 

Although all calculations are based exclusively on the reduced data (1) and the crystal structure (2), and thus, diffraction images (3) are not strictly necessary for \textsc{Garfield} to work, it is strongly recommended to attach representative diffraction image files to a project (see section \ref{sec:design}). Full functionality is only attained if a diffraction image has been assigned. For instance, if a diffraction image is assigned, it is also possible to define a mask in order to exclude data from unreliable areas of the detector, such as the shadow of a beamstop. Furthermore, the quality of the results can be easily cross-checked within \textsc{Garfield} by comparing measured with simulated diffraction images. Otherwise, visual comparison is only possible by means of \textit{diffraction patterns} (in \textsc{Garfield}'s parlance), i.e. by two-dimensional scatter plots showing the positions and intensities of (observed or calculated) Bragg spots.  

In \textsc{Garfield}'s terminology, which is also adopted in the accompanying documentation and the remainder of this report, a diffraction \emph{pattern} is a two-dimensional graphical representation of the reduced data (`obs') or the predicted version of them (`clc'), obtained by plotting circular disks of different size and gray level (to indicate the intensities $I_{i}^{\mathrm{foo}}$) centered at coordinates $(x_{i}^{\mathrm{foo}}, y_{i}^{\mathrm{foo}})$,  `foo' = `obs' or `clc'. In contrast, a diffraction \emph{image} corresponds to a matrix of gray values representing the pixel-by-pixel distribution of diffraction intensities, as if recorded with a two-dimensional array detector. \textsc{Garfield} provides options to view the observed diffraction images or patterns, and to compare them with simulated counterparts.

The main window of \textsc{Garfield} displays a list of parameter sets ("settings") for simulating the diffraction data of the current project (see Figure S1 in the supplementary material).
If the current project has just been created, the list consists of a single line, representing a dummy parameter set. In this case, in order to start working with the new project, the user must first create a new parameter set by editing the default parameters and assigning them realistic or at least meaningful values. Most of the parameters (including the orientation of the crystal lattice) should be reasonably well known from the experimental setup, but sometimes this is not the case. Other parameters, such as the mosaicity and the average size and shape of the crystallites, are usually not known in advance and must be fitted to the data by using the \textsc{GeoFit} tool.                                                                                                 \subsection{The main window} \label{sec:work.3}

The main window is used to list and manage a growing collection of parameter sets used to explore the parameter space in order to find the optimal description of the diffraction experiment. New parameter sets can be added by manually editing existing sets or by saving the results of \textsc{GridScan} and \textsc{GeoFit} runs. 

A complete set consists of model parameters (Table \ref{tab:Table_I}) and control parameters (Table \ref{tab:Table_II}). The former describe the diffraction geometry and sample properties.
Many of them can be fitted with \textsc{GeoFit}. Control parameters include binary flags and numerical parameters that define details of how diffraction images are predicted and fitted, such as the type of modeling to be used ("standard" or "ANISO", see section \ref{sec:model.4}), the resolution range, or the \textit{ambit radius} (radius of circular areas around observed diffraction spots used in the definition of the cost function, see section \ref{sec:model.1}), and whether all predictions or only matching ones ("hits") should be considered (for further explanations see section \ref{sec:model}).   

\begin{table}
\caption{\label{tab:Table_I}Model parameters}
\begin{ruledtabular}
\begin{tabular}{lccl}
Parameter(s) & edit\footnotemark[1]&fit\footnotemark[2] & Comment\\
\hline
Electron energy & \checkmark & & in keV\\
Wavelength & \checkmark & & $\lambda$ \\
Divergence & \checkmark & & half-angle of circular cone\\
Bandwidth & \checkmark & & $\Delta\lambda/\lambda$ \\
Pixel size & \checkmark & & equal in X and Y (LAB coord.s)\footnotemark[3]\\
Detector distance & \checkmark & & $L$, real distance sample$\--$detector\\
Camera length & \checkmark & & $L_{\mathrm{eff}}$\\
Magnification & \checkmark & \checkmark & $L_{\mathrm{eff}} / L$ \\
Beam center (X, Y)& \checkmark & \checkmark & on the detector (in pixels)\\
Image rotation\footnotemark[4] & \checkmark & (\checkmark) & due to magnetic lens\\
Image distortion & \checkmark & \checkmark & elliptical correction (2 param.s)\\
Orientation & \checkmark & \checkmark & of crystal lattice (3 param's)\\
Mosaicity & \checkmark & \checkmark & orientation spread of domains\footnotemark[5]\\
Shape transform & \checkmark & \checkmark & approximated by Gauss function\footnotemark[6]\\
Scale factor & & \checkmark & overall intensity scaling\\
B factor & & \checkmark & overall B factor correction\\
Omega axis\footnotemark[7] & (\checkmark) & & direction of tilt axis\footnotemark[8]\\
Scale corrections\footnotemark[7] & & (\checkmark) & $n - 1 $ correc. factors for $n$ images\\
Omega corrections\footnotemark[7] & & (\checkmark) & $n - 1 $ angles for $n$ images\\
\end{tabular}
\end{ruledtabular}
\footnotetext[1]{Parameters to be set in the Edit menu; redundant values are updated automatically.}
\footnotetext[2]{Parameters that can be fitted with \textsc{GeoFit}.}
\footnotetext[3]{X, Y, Z: Cartesian coordinate system, Z in direction of the electron beam.}
\footnotetext[4]{Cannot be fitted in a single-image project.}
\footnotetext[5]{Standard model: normal distribution (1 parameter); ANISO model: covariance matrix (6 parameters). }
\footnotetext[6]{ Approximation of the squared magnitude of the effective shape transform.
Standard model: width ($\sigma$) and direction (tilt from Z) of reciprocal lattice rods (3 parameters); ANISO model: covariance matrix of a 3D normal function (6 parameters).}
\footnotetext[7]{Only available in multi-image projects (tilt series).}
\footnotetext[8]{The tilt axis must be perpendicular to Z (1 parameter).}
\end{table}

\begin{table}
\caption{\label{tab:Table_II}Control parameters}
\begin{ruledtabular}
\begin{tabular}{ll}
Parameter & Description \\
\hline
\texttt{dmin, dmax} & resolution range (upper and lower d-spacing in Å)  \\
\texttt{ambit} & radius (in pixels) of the spot ambits \\ 
\texttt{mask} & boolean: apply masking of bad detector regions \\
\texttt{target} & flag: "I" or "sqrt(I), see  equ. (\ref{eq8})\\
\texttt{Pos.weight} & weight of position errors, see  equ. (\ref{eq1})\\
\texttt{fPred.weight} & wt of non-matching predictions, $= w_2$ in equ. (\ref{eq7}) \\
\texttt{ANISO} & boolean: ANISO (true) or standard model (false) \\
\texttt{Tilt.use\footnotemark[1]} & boolean: if true, take tilt of relrods into account \\
\texttt{K0.corr\footnotemark[2]} & boolean: if true apply per-tilt intensity corrections \\
\texttt{Omg.corr\footnotemark[2]} & boolean: if true apply per-tilt angle corrections \\
\end{tabular}
\end{ruledtabular}
\footnotetext[1]{Only used in standard modeling.}
\footnotetext[2]{Only available in multi-image projects (tilt series).}
\end{table}

Any setting can be selected as the "active setting" by double-clicking on a row in the main table. Only one setting can be "active" at a time. Most of the options available via drop-down menus can only be used with the active setting defined. The parameters of the active setting are used when simulating diffraction patterns and diffraction images (menu "Display"), when searching for possible crystal orientations (menu "GridScan"), or as starting values for NLS parameter fitting (menu "GeoFit").  

\subsection{The GridScan tool} \label{sec:work.4}

\begin{figure*}[ht]
\includegraphics[width=\textwidth]{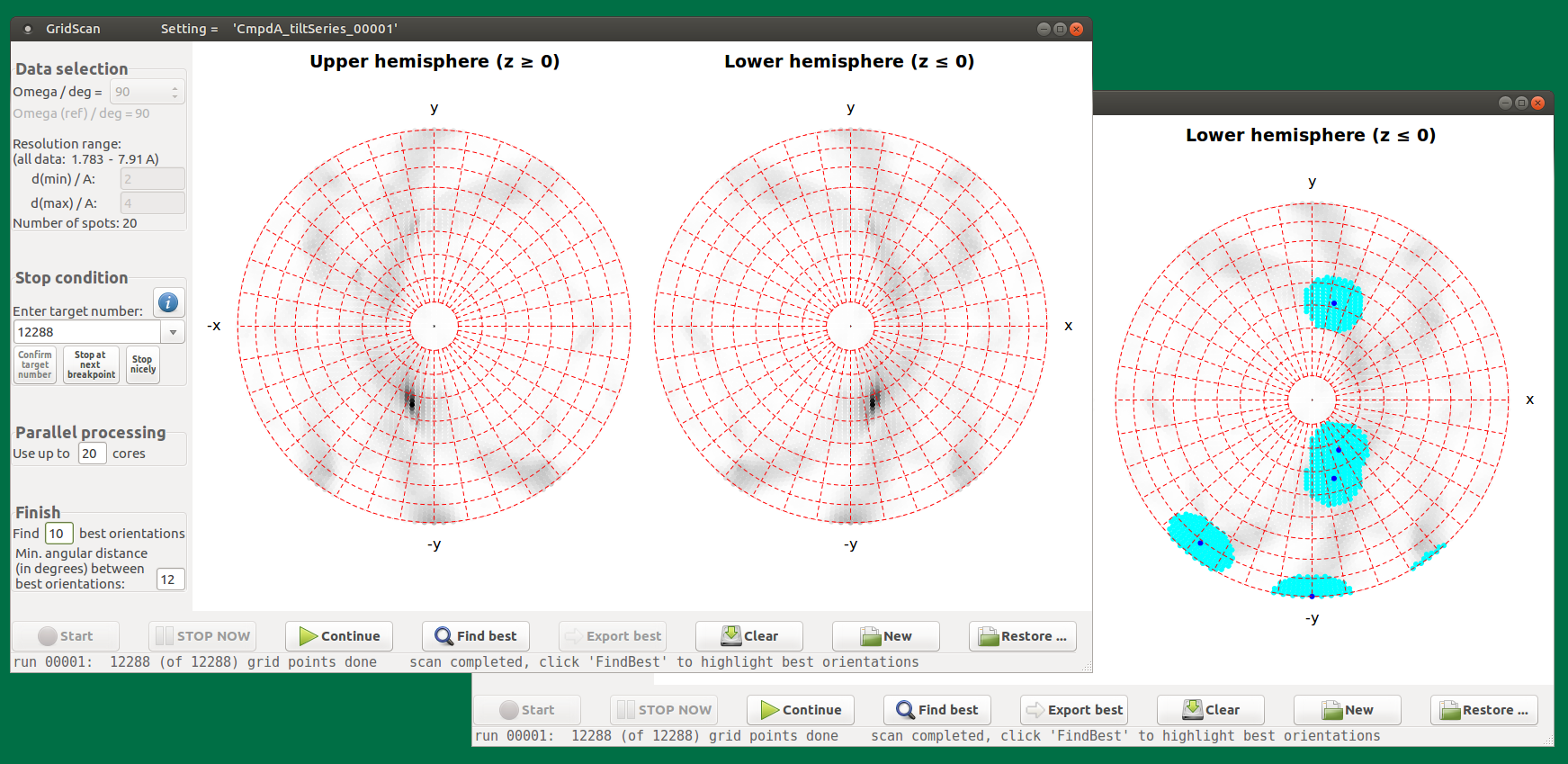}
\caption{The \textsc{GridScan} window. All calculations in \textsc{GridScan} are performed with the parameters of the active setting, except for the resolution limits, d(min) and d(max), and the crystal orientation parameters, which are systematically varied so that the sampling of all possible orientations is as uniform as possible. The orientations are parameterized by the direction of the incident beam relative to the (fixed) crystal coordinate system, followed by a rotation of the crystal around the beam. A figure of merit (FOM) is calculated for each orientation, which measures the quality of the match between the predicted and observed diffraction pattern. The polar plots in the main part of the window show the maximum FOM found for each direction of the beam as density maps in equal-area Lambert projections of the upper and lower hemispheres. On the left margin, data used for \textsc{GridScan} can be filtered by restricting the resolution range. (For a multi-image project, one image must be selected.) The "resolution" (mesh size) of the grid to be searched is determined by the number of grid points tested ("target number"). Any number up to 49152 (directions of the incident beam) can be entered. The computation can be stopped at any time (and continued later, if desired). Parallel processing is possible if multiple CPU cores are available. \textbf{Foreground}: FOM maps after testing 12288 directions of the incident beam. \textbf{Background}: same as in foreground panel, with the 10 directions of highest FOM  that are at least 12 degrees apart from each other marked with a black dot in a cyan-colored surrounding.}
\label{fig:CmpdA_gridscan}
\end{figure*}

Once a new setting with more realistic parameter values has been created by manual editing, it is advisable to compare the diffraction pattern predicted by this setting with the observed pattern. If the predicted pattern shows resemblance to the observed pattern, it might be possible to proceed directly to the final step of optimizing the parameters by NLS fitting with \textsc{GeoFit}. Often this is not the case, and for the parameters that are not known precisely enough, better start values for \textsc{GeoFit} must be found. The most critical parameters are clearly the crystal orientation parameters, because even a small misalignment can completely change the diffraction pattern. Other parameters that are typically affected by large uncertainties, such as mosaicity and effective domain size, are less critical, because the diffraction changes continuously with these parameters.
\footnote{At the beginning of a project, when the orientation of the crystal is not yet established, overestimating the mosaicity can actually be beneficial for NLS fitting.}
  
The \textsc{GridScan} tool is an indispensable aid for finding better candidates for the  orientation of the crystal lattice by performing a grid search over the entire $\mathsf{SO(3)}$ rotation space.
\footnote{\textsc{GridScan} does by design not exploit crystal symmetry to minimize the region of three-dimensional rotation space to be scanned. Thus, potential complications due to improper rotations (which change the handedness) are excluded. On the other hand, symmetry-equivalent orientations should appear in the map created by \textsc{GridScan} as sets of candidate orientations with similar figures of merit.} 
When the tool is activated, a new window opens for setting up, starting, stopping, and monitoring the progress of grid searches (Figure \ref{fig:CmpdA_gridscan}). In each run, the parameters of the active setting (the one that was active when \textsc{GridScan} was launched), except the orientation parameters and the resolution range, are used to compute a figure of merit (FOM) for up to 17,694,720 ($= 49152 \times 360$) orientations, by comparing the predicted patterns with the "observed" pattern. 

In \textsc{Garfield}, the orientation of the crystal is defined by two polar angles, $\Theta$ and $\Phi$, which define the direction of the incident beam relative to the crystal lattice, and a third angle, $\Psi$, for the rotation of the crystal around the incident beam.
\footnote{Note that ($\Theta, \Phi, \Psi$) are not Euler angles.}
The grid is constructed from 49152 predefined directions of the beam, $(\Theta, \Phi)$, each one combined with 360 rotations sampling $\Psi$ in steps of one degree. The beam directions and the order in which they are visited are defined according to the principles of HEALPix (Hierarchical Equal Area isoLatitude Pixelation\cite{Gorski2005}). This sampling results in a most uniform and efficient
\footnote{HEALPix sampling of the unit sphere is most efficient in the sense that sampling with a set of $n$ grid points is the most uniform sampling possible with $n$ points provided that the same applies for sampling with any larger set 
obtained by adding more grid points.} 
coverage of the sphere with an angular resolution of 0.92°. The angular distance of any arbitrary orientation from the nearest grid point is less than one degree.

For each pair $(\Theta, \Phi)$, represented by a point on the unit sphere, the maximum value of the FOM obtained by scanning over $\Psi$ is plotted in two-dimensional gray-scale maps showing the projections of the upper and lower hemispheres onto the $x,y$ plane
\footnote{A Cartesian coordinate system with basis vectors $\vec{e}_x, \vec{e}_y, \vec{e}_z$ fixed in the crystal lattice is defined by $\vec{a} \parallel \vec{e}_x$ and ${\vec{c}}^{\star} \parallel \vec{e}_z$.}.
Good candidate orientations will show up as dark spots in these maps. On a modern multi-core workstation, a full grid scan takes only a few minutes to half an hour, depending on the number of diffraction spots used (typically 50 to 100). This number depends on the resolution range used for the grid scan, which can be set in the \textsc{GridScan} window before starting a scan. Also the number $N$ of beam directions to evaluate has to be set before starting a run ($N \le 49152$). A running grid scan can be aborted at any time (and resumed or extended at a later time, if this seems promising). Once a grid scan is finished (or has been terminated), the $n$ best orientations on the FOM ranking can be exported to \textsc{Garfield}'s main table, creating $n$ new settings.

\subsection{The GeoFit tool} \label{sec:work.5}

After creating a new setting (either by manual editing or using \textsc{GridScan}), it is usually necessary to optimize the current parameters through NLS fitting using the \textsc{GeoFit} tool (for more details, see supplementary material, Figure S2). The last step in a project should always be to run \textsc{GeoFit} to optimize the parameters and get a summary of the final results.

\begin{figure*}[htbp]
\includegraphics[width=\textwidth]{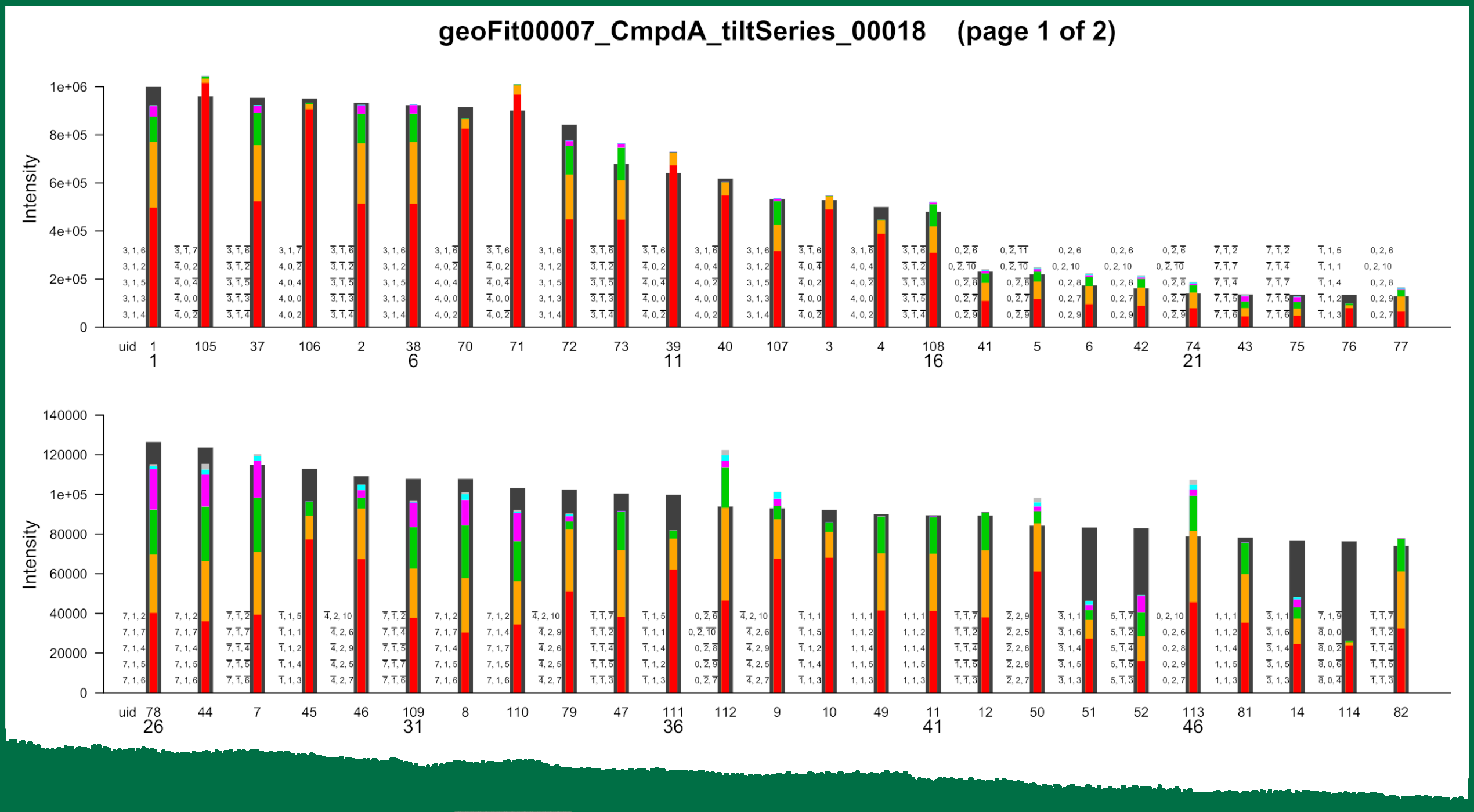}
\caption{ Observed spot intensities and the contributions of individual Bragg reflections according to \textsc{GeoFit}. Top part of the first page of the graphical output (pdf file) created with \textsc{GeoFit} run 00007 of project "CmpdA-tiltSeries", starting from the parameters of setting 00018. The spots are sorted in descending order of observed intensities. The observed intensities are represented by black bars, and the calculated values by colored bars in front of the black bars (mostly obscuring the black bars). The length of the colored bars is divided in up to 6 segments proportional to the relative contributions of up to 5 individual Bragg reflections. The contributions are sorted from strongest (bottom) to weakest (top). If more than 5 reflections are contributing, a sixth segment (gray) representing the sum of the remaining contributions appears on top of the other segments. The Laue indices of the 5 strongest reflections are written next to the bars in the same order from bottom to top.}
\label{fig:CmpdA_barDiagr}
\end{figure*}

The fitted values of all model parameters and the control parameters used for fitting can be saved for further use. Saving the results creates a new setting and adds a new entry to the \textsc{Garfield}'s main table. In addition, several files are written to the file system for documentation and export, including (1) a listing of all parameters with their values before and after fitting, (2) a corresponding list of various kinds of SSR values and R factors, and (3) a comprehensive table of data related to Bragg reflections, combining the input coordinates and intensities of observed Bragg spots, the corresponding calculated intensities and centroid positions, and \textendash~for each spot \textendash~the proposed reflection indices $hkl$, expected intensities, and excitation errors (distances from the Ewald sphere) of up to five Bragg reflections contributing to the Bragg spot. A pdf file with a graphical representation of the most relevant information contained in the table (3) is also generated (see Figure \ref{fig:CmpdA_barDiagr}).

Note that in the example of Figure \ref{fig:CmpdA_barDiagr} none of the observed spots is caused by a single Bragg reflection, but all spots are superpositions of multiple reflections. By considering spots with a dominant and one or several minor contributions, two extreme cases can be distinguished: (i) The structure factors are comparable in magnitude, but the excitation errors of the minor contributions are large compared to the dominant reflection. (ii) The excitation errors are comparable, but the structure factors are very different in magnitude. In case (i), but not in case (ii), it may be justified to neglect the contribution of spurious reflections to the temporal intensity variations measured in UED experiments. 

\section{Prediction of diffraction patterns}\label{sec:model}

The purpose of the two core tools of \textsc{Garfield}, \textsc{GridScan} and \textsc{GeoFit}, is to find model parameters that reproduce the input data, the positions and intensities of recorded Bragg spots (the \textit{reduced data}). Bragg spots are either individual Bragg reflections or superpositions of them. In \textsc{Garfield}'s terminology, the term \textit{diffraction pattern} refers to the graphical representation of the \textit{reduced data}. Thus, the two terms are essentially equivalent.

Predicting a diffraction pattern for comparison with the observed pattern requires (1) a crystallographic model describing the diffraction by the sample, and suitable algorithms to (2) estimate the expected diffraction and to (3) convert the result into a list of triplets $(x_{i_\mathrm{o}}^{\mathrm{clc}}, y_{i_\mathrm{o}}^{\mathrm{clc}}, I_{i_\mathrm{o}}^{\mathrm{clc}})$ analogous to the input list of reduced data $(x_{i_\mathrm{o}}^{\mathrm{obs}}, y_{i_\mathrm{o}}^{\mathrm{obs}}, I_{i_\mathrm{o}}^{\mathrm{obs}})$, where index $i_\mathrm{o} = 1, ...~N_\mathrm{obs}$ enumerates the observed Bragg spots. In order to be of practical value for NLS fitting of the model parameters, the model must be simple to allow fast computation. Also, the focus is on diffraction \textit{patterns} (rather than diffraction \textit{images}) because the algorithms computing the cost function in NLS fitting should be as fast as possible. 

The potential loss of precision of the predicted spot positions $(x_{i_\mathrm{o}}^{\mathrm{clc}}, y_{i_\mathrm{o}}^{\mathrm{clc}})$ due to various simplifications and approximations in the model is compensated by additional restraints imposed on the model by considering the spot intensities as input data.  The goal is to find the solution corresponding to the global minimum of the cost function. Taking intensities into account increases the contrast between different solutions, even if the calculated intensities $I_{i_\mathrm{o}}^{\mathrm{clc}}$ suffer from large uncertainties, but the correlation between estimated and observed intensities is positive. If a well separated solution can be identified, the orientation of the crystal lattice is usually quite robust to errors in the calculated intensities, so that the assignment of Laue indices is not affected. Thus, the effect of considering intensities can be understood as the application of a filter that eliminates unlikely solutions. The effect of intensity uncertainties is to reduce the selectivity of the filter, rather than to bias the solution toward a completely different set of parameters, i.e. a solution that would produce a wrong indexing.

The following subsections describe how the above requirements (1--3) are implemented in \textsc{Garfield}. We start with (3), explaining how the predicted diffraction pattern is derived from the simulated diffraction. This provides an opportunity to introduce the concept of \textit{spot ambits}, which is essential for understanding the whole approach and helps to put the limitations of the diffraction model into the right perspective. This first subsection (A) is followed by the definition of the cost function used in NLS fitting (B). The last two subsections describe the model itself (C) and how it is used to simulate the diffraction (D).

\subsection{From the diffraction to the diffraction pattern} \label{sec:model.1}

In the kinematical theory of diffraction, a perfect crystal in a parallel, monochromatic beam will diffract only in well-defined directions and only if the crystal is oriented so that at least one point of the reciprocal lattice (in addition to the origin) fulfils the Bragg condition by intersecting with the Ewald sphere. If the crystal consists of small domains with slightly different orientations and the beam is not perfectly coherent, RLPs near the Ewald sphere can also give rise to diffraction. This can be viewed as an expansion of the RLPs from points of zero extent to distributions over a finite volume. When treated as distributions in reciprocal space, virtually all RLPs (up to a certain resolution) can contribute to diffraction, but most of them, however, with an intensity that is zero or near-zero.

For now, let's assume that the contributions of all RLPs, enumerated by their Laue indices $hkl$, have been calculated (as described in subsections \ref{sec:model.3} and \ref{sec:model.4}) and are available as a table of centroid coordinates, integrated intensities, and corresponding Laue indices:  $x_{j_\mathrm{p}},\, y_{j_\mathrm{p}},\, I_{j_\mathrm{p}}$, and $(hkl)_{j_\mathrm{p}}$ with $j_\mathrm{p} = 1, ...~N_\mathrm{p}$. The number $N_\mathrm{p}$ of predicted reflections is usually much larger than the number of observed spots $N_\mathrm{obs}$. 

Since both the observed spot positions and the calculated predictions may be subject to ambiguity and uncertainty, a circular disk in the detector plane
\footnote{The detector is assumed to be perpendicular to the incident beam, which travels along the $z$ axis.}
around $(x_{i_\mathrm{o}}^\mathrm{obs}, y_{i_\mathrm{o}}^\mathrm{obs})$ is defined as a neighbourhood $A_{i_\mathrm{o}}$ for each observation $i_\mathrm{o}$, 
\begin{equation} \label{eq2}
A_{i_\mathrm{o}} = \left\{ (x,y) ~ \big| ~ (x - x_{i_\mathrm{o}}^\mathrm{obs})^2 + (y - y_{i_\mathrm{o}}^\mathrm{obs})^2 < r_\mathrm{a}^2 \right\},
\end{equation}
and predictions $j_\mathrm{p}$ are taken as contributions to spot $i_\mathrm{o}$ if their centroids are within the spot's neighbourhood, $(x_{j_\mathrm{p}},\, y_{j_\mathrm{p}}) \in A_{i_\mathrm{o}}$. The disks $A_{i_\mathrm{o}}$ of radius $r_\mathrm{a}$ will be referred to as the \emph{ambits} 
\footnote{Note that the ambit radius $r_\mathrm{a}$ is the same for all observations. Normally the default value should not need to be changed. If the default radius is larger than necessary, it can easily be decreased. It can likewise also be increased, with the consequence that spots with overlapping ambits are  automatically merged. The user has to make sure that the newly defined observations and their ambits make sense after merging. This is facilitated by the interactivity of the GUI.} of the observed spots. By default, $r_\mathrm{a}$ is the maximum possible radius with no overlapping ambits, such that any point  is either element of exactly one ambit or not element of any $A_{i_\mathrm{o}}$.

Thus, each prediction $j_\mathrm{p}$ is assigned to exactly one observed spot $i_\mathrm{o}$ or none. For convenience, we define an index set $J = \left\{1, ...\,N_\mathrm{p} \right\}$ and disjoint subsets of $J$,
\begin{equation} \label{eq3}
J_{i_\mathrm{o}} =   \left\{ j_\mathrm{p} \in J ~ \Big| ~ (x_{j_\mathrm{p}},\, y_{j_\mathrm{p}}) \in A_{i_\mathrm{o}}  \right\} \qquad \forall{ i_\mathrm{o} \in \left\{1, ...\,N_\mathrm{obs}\right\} },
\end{equation}
as well as a set $J^\mathrm{nohits}$ pointing to the predictions that do not fall within the ambit of any observed spot:
\begin{equation} \label{eq4}
J^\mathrm{nohits} = J \setminus \bigcup_{i_\mathrm{o}} J_{i_\mathrm{o}}
\end{equation}

Combining all predictions contributing to a given observation means addition of the intensities and taking the weighted average of the centroid coordinates. Thus,
\begin{align} 
I_{i_\mathrm{o}}^{\mathrm{clc}} &= \sum_{j_\mathrm{p} \in J_{i_\mathrm{o}}} I_{j_\mathrm{p}} \label{eq5}
\\
x_{i_\mathrm{o}}^{\mathrm{clc}} &= ({I_{i_\mathrm{o}}^{\mathrm{clc}}})^{-1}  \sum_{j_\mathrm{p} \in J_{i_\mathrm{o}}} x_{j_\mathrm{p}} ~ I_{j_\mathrm{p}}, \quad \text{if} ~ J_{i_\mathrm{o}} \ne \emptyset, \quad
(y_{i_\mathrm{o}}^{\mathrm{clc}} ~ \mathrm{alike}) \label{eq6} 
\end{align}

For quick comparison with the reduced data (the input diffraction pattern), \textsc{Garfield} provides the option to calculate the values in equations \ref{eq5} and \ref{eq6} for any setting in the main table and to display the result as simulated diffraction pattern.

\subsection{The cost function}\label{sec:model.2}

The parameters of a selected setting can be fitted with the \textsc{GeoFit} tool by NLS minimization of a cost function $S$, which is defined as the sum of squared differences between calculated and observed quantities that are related to the intensities and coordinates of diffraction spots (cf. equation \ref{eq1}). The exact form of the cost function can be adjusted by using a set of control parameters. One of the control parameters is $w_\mathrm{P}$ already introduced in equation \ref{eq1}. Other control parameters are the minimum and maximum resolution of the diffraction spots to be considered.
\footnote{The resolution range and the possibility to exclude certain areas of the detector plane by applying a mask will not be mentioned further below in order not to overload the formal description.}

Two control parameters that should be mentioned are the flag $\mathsf{target}$ and another weighting factor, $w_2$. The first takes character string values "$\mathsf{I}$" or "$\mathsf{sqrt(I)}$" and determines whether the residuals based on intensities are calculated from the intensities directly or from the square roots of intensities. Essentially, this allows the user to change the weighting scheme used for the intensity dependent part $S^\mathrm{(I)}$ of the cost function. The second, $w_2$, determines whether predicted reflections that are not within the ambit of any observation (most often "false predictions") are simply ignored ($w_2 = 0$, the default), or fitted towards zero intensity ($w_2 > 0$). Positive values of $w_2$ should be used with caution, and only when \textsc{GeoFit} is trapped in a local minimum and tries to improve the fit of the observed intensities by generating many false reflections outside the ambits of all observed spots.
Thus, 
\begin{equation} \label{eq7}
S ~ = ~ S^\mathrm{(I,hits)} ~  + ~ w_2 \, S^\mathrm{(I,nohits)}~ + ~ w_\mathrm{P} \, S^\mathrm{(P)}
\end{equation} 
where the intensity part $S^{(I)}$ has been split into two contributions.
\begin{equation} \label{eq8}
S^\mathrm{(I,hits)} = \begin{cases}
\sum_{i_\mathrm{o}} w_{i_\mathrm{o}} (I_{i_\mathrm{o}}^{\mathrm{clc}} - I_{i_\mathrm{o}}^{\mathrm{obs}})^2, 
& \text{if } \mathsf{target} = \mathsf{"I"}  \\ 
\\  
\sum_{i_\mathrm{o}} \big(\sqrt{I_{i_\mathrm{o}}^{\mathrm{clc}}} - \sqrt{I_{i_\mathrm{o}}^{\mathrm{obs}}}\big)^2,
& \text{if } \mathsf{target} = \mathsf{"sqrt(I)"}.
\end{cases}
\end{equation} 

If $w_2 > 0$, the contribution to $S^\mathrm{(I)}$ by the predictions that cannot be assigned to any observation is calculated analogously, by setting the corresponding $I^\mathrm{obs}$ values to zero.
\footnote{Weight $w_2$ should not be set to a positive value unless the list of reduced data is complete, i.e. contains all observations that are theoretically possible. If this is not the case, e.g. because reflections are hidden behind the beamstop, a suitable mask should be defined.}
\begin{equation} \label{eq9}
S^\mathrm{(I,nohits)} = \begin{cases}
\displaystyle{\sum_{j_\mathrm{p} \in J^\mathrm{nohits}} (I_{j_\mathrm{p}}^{\mathrm{clc}})^2}, 
&  \quad\quad \, \text{if } \mathsf{target} = \mathsf{"I"}  \quad \\ 
\\  
\displaystyle{\sum_{j_\mathrm{p} \in J^\mathrm{nohits}} I_{j_\mathrm{p}}^{\mathrm{clc}}} ,
&  \quad\quad \, \text{if } \mathsf{target} = \mathsf{"sqrt(I)"}. \quad
\end{cases}
\end{equation} 

The $w_{i_\mathrm{o}}$ in equation \ref{eq8} are individual weights, which in principle should be read from the input list of reduced data. However, weights (or sigma values) assigned to observed intensities are not taken into account in the present version of \textsc{Garfield}. Instead, the weights are determined as $w_{i_\mathrm{o}} = (I_{i_\mathrm{o}}^\mathrm{obs})^{-\frac{1}{2}}$, which has the effect to weighing down high intensities relative to low intensities, similar to the effect of setting $\mathsf{target} = \mathsf{"sqrt(I)"}$.

The position part of the cost function is defined as the sum of squared distances between predicted and observed spot positions, $d_{i_\mathrm{o}}^2 = (x_{i_\mathrm{o}}^\mathrm{clc} - x_{i_\mathrm{o}}^\mathrm{obs})^2 + (y_{i_\mathrm{o}}^\mathrm{clc} - y_{i_\mathrm{o}}^\mathrm{obs})^2 ~ \text{if} ~ J_{i_\mathrm{o}} \ne \emptyset, ~ r_a^2 ~ \text{otherwise}$. Thus,
\begin{equation} \label{eq10}
S^\mathrm{(P)} = \sum_{i_\mathrm{o}} d_{i_\mathrm{o}}^2.
\end{equation} 

It should be noted that position information present in the reduced data enters the cost function via the ambits, even if $w_\mathrm{P} = 0$ (the default) and, thus, $S^\mathrm{(P)} = 0$.

\subsection{Electron diffraction by typical "UED crystals"}\label{sec:model.3}

All simulations in \textsc{Garfield} are calculated within the kinematical theory of diffraction. The use of dynamical theory, which should be better suited for electron diffraction, is impractical because the computational effort would be too high. Furthermore, most samples used in UED experiments are far from perfect crystals due to the thickness constraints and therefore do not lend themselves to a rigorous treatment of multiple scattering. Thus, using a kinematical approximation is the only practicable way for the present purpose. Fortunately, the restriction to thin crystals may not be as severe as it seems, since rather large errors in the estimated diffraction intensities can be tolerated, as outlined at the beginning of section \ref{sec:work}. In addition, there are reasons to belief that dynamical effects are less pronounced in highly textured crystal, i.e. when the crystallites are small and the mosaic spread is large\cite{Cowley2006a, Dorset2010} as is often the case due to typical limitations of UED experiments. 

Ignoring dynamical diffraction effects reduces the description of electron diffraction to a mere problem of considering the relative phases of waves scattered by the distribution of matter in the crystal, in strict analogy to X-ray diffraction. This leads to the Bragg or Laue conditions of diffraction, which can be visualized with the Ewald sphere construction. The transition from X-ray to electron diffraction is then essentially a matter of replacing X-ray atomic scattering factors with electron scattering factors and taking into account that for electrons, the radius of the Ewald sphere, $1/\lambda$, is much larger than for X-rays and almost "plane-like". \textsc{Garfield} uses the parameterizations of electron scattering factors of atoms and ions by Peng et al. \cite{Peng1996, Peng1998}

The restriction to very thin samples relaxes the diffraction conditions, to some extent, so that points of the reciprocal lattice can give rise to reflections even if the Laue conditions are not exactly fulfilled, provided that the \textit{excitation error}, the distance of an RLP from the Ewald sphere, is not too large. Usually, this situation is interpreted in a different way: instead of relaxing the Laue conditions, the RLPs are replaced by needle-shaped distributions parallel to the surface normal of the crystal sheet, and diffraction occurs, if a "reciprocal lattice rod" (\textit{relrod}) intersects with the Ewald sphere. This is consistent with the idea, that the reciprocal lattice weighted with the structure factor represents the Fourier transform of a perfect crystal of infinite size, $\mathcal{F}_\infty$, whereas the Fourier transform of a finite crystal is the convolution of $\mathcal{F}_\infty$ with the Fourier transform of the crystal's shape function. For a plane parallel plate of thickness $t$ perpendicular to the incident beam, this leads to relrod profiles oscillating like a $\mathit{sinc}$ function. The intensity measured for a particular RLP with indices $hkl$ varies as $\mathit{sinc}^2(\pi \, t \, \Delta h)$ with the distance $\Delta h$ of the RLP from the Ewald sphere (measured in units of $h = 1/d_{hkl}$, the inverse of the d-spacing).

In UED experiments, the finite size of the crystal is not the only, and often not the most important reason why non-zero diffraction intensities occur for RLPs even if they are nominally not located on the Ewald sphere. Because of the large area of the diffracting volume (compared to its thickness) and the non-perfect quality of most samples, the orientation of the crystal lattice is not exactly the same in different parts of the sample. Thus, UED samples should rather be viewed as an ensemble of crystalline domains with slightly different orientations distributed around the average orientation of the crystal lattice. This is similar to the orientation distribution of mosaic blocks introduced by Darwin \cite{Darwin1922} to quantitatively explain X-ray diffraction of single crystals. Although the orientation distribution in UED experiments can be much wider (typically several degrees) than the typical mosaic spread of good single crystals (usually $\ll 0.1^{\circ}$), the effect of orientation disorder will be referred to as \textit{mosaicity} in the following. 
Other effects that contribute to the appearance of reflection spots in UED images are the divergence of the incident beam and, to a lesser extent, its finite energy bandwidth. In \textsc{Garfield}, all these effects are accounted for by following the example of the relrods introduced to explain the finite crystal thickness effect: RLPs (represented by Dirac delta functions) are replaced by smooth distributions, which should lead to correct results, at least approximately, if the Ewald sphere construction is used without modification. 

To allow simple and fast computation of the combined effect, \textsc{Garfield} assumes normal distributions to describe the four effects. Gaussian approximations have been used in various formulations in serial snapshot crystallography to estimate reflection intensities and partialities (see Brehm et al., 2023\nocite{Brehm2023}\cite{Brehm2023} and the references cited therein). In \textsc{Garfield}, the simplest case is assumed: each individual effect is described by a single Gaussian distribution, and the distributions are mutually independent. Then, the combination of any two effects corresponds to the convolution of two Gaussians, which in turn is a normal distribution. Furthermore, the marginal and conditional distributions of multivariate Gaussians are normal distributions. This allows a simple analytical formula to be used in many cases, while more realistic distributions would require time-consuming numerical integration. A few remarks are in order for all four effects.

\subsubsection{Finite size effect}

\begin{figure}[ht]
\centering
\includegraphics[width=8.5cm]{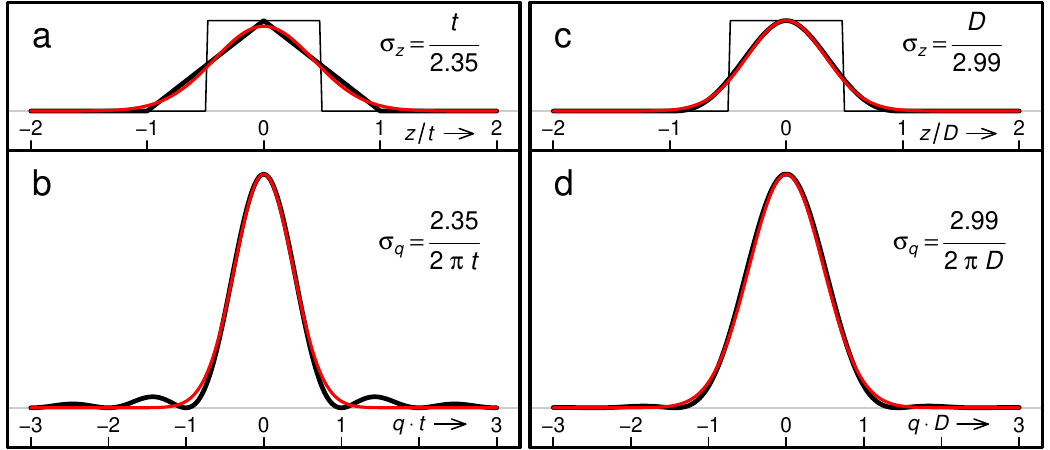}
\caption{Autocorrelation function (ACF) and the intensity profile ($P_\mathrm{I}$) of the RLPs for a thin plate (\textbf{a}, \textbf{b}) and a solid sphere (\textbf{c}, \textbf{d}), and their approximation by normal distributions (red curves).
\textbf{(a)}: volume function (box-like function) and ACF (thick black line) along real space coordinate $z$ perpendicular to a thin plate (thickness $t \ll $ lateral extent); \textbf{(b)}: $P_\mathrm{I}$, the Fourier transform of the ACF shown in (a), and the best approximation of $P_\mathrm{I}$ by a normal distribution of the same height; \textbf{(c)}: similar to (a) for a solid sphere of diameter $D$, except that the thick black line represents the \textit{projection} of the ACF to the $z$ direction (arbitrary); \textbf{(d)}:  Fourier transform of the ACF projection shown in (c) and best approximation by a normal function. $\sigma_z$ and $\sigma_q$ are the standard deviations of the normal distributions in real and reciprocal space, respectively. The numerical factors (2.35 for a thin plate, 2.99 for a sphere) were determined by minimizing the maximum of the absolute differences between $P_\mathrm{I}$ and its approximation.}
\label{fig:ShapeFunctions}
\end{figure}

Given the fact that the intensity profile of the relrods for a thin plate (Fig. \ref{fig:ShapeFunctions}b) oscillates like the square of the $\mathit{sinc}$ function, using normal distributions to describe the finite size effect may seem unphysical. However, the set of coherently diffracting domains (crystallites) of a typical UED sample varies not only in orientation but also in lateral extent and thickness. Averaging over crystallites of different thicknesses will fill in the minima of the intensity profile, $P_\mathrm{I}(z)$, and results in smoothing, making the profile sufficiently normal. The smoothing will be even more pronounced if limitation of the domains in lateral directions (perpendicular to the beam) cannot be neglected. In this case, the relrods must be described by three-dimensional distributions of finite extension perpendicular to their main direction. An extreme case of diffraction by isometric particles is represented in Figure \ref{fig:ShapeFunctions}d by the intensity profile for the solid sphere.

The intensity profile of the RLPs due to the finite size effect is described by a general three-dimensional normal distribution with 6 independent parameters: three standard deviations $\sigma_i^\mathrm{shp}$ along the principal axes ($i = 1,2,3$) and three Euler angles. Degenerate distributions (one or two of the $\sigma_i^\mathrm{shp}$ equal to zero) are allowed. \textsc{Garfield} is able to handle the case of a thin plate, all other broadening effects being negligible. However, due to the restriction to Gaussian profiles, the NLS fitting of thickness $t$ is biased towards smaller values than the true average $\bar{t}$, since in this way the slow decay of the side maxima of the $\mathit{sinc}$ function can be compensated by increasing the width of the Gaussian profile, $\sigma_q \propto t^{-1}$. In other words, Garfield cannot be used to fit the geometric thickness of the sample because the parameter t is correlated with, but not identical to, an effective thickness.

\subsubsection{Mosaicity}

In contrast to the finite size effect, which is the same for all RLPs, the broadening of RLPs due to the orientation distribution of mosaic blocks
\footnote{Here, coherently diffracting domains or \textit{mosaic blocks} are considered to be small but otherwise perfect crystals. Variation in cell constants are not taken into account. This is in line with the original treatment of mosaicity by Darwin (1922) \cite{Darwin1922}.}
increases linearly with the resolution, $q = 1/d$. Here, $d$ is the "d-spacing", $q = 2 \sin\theta / \lambda$, $\theta$ the Bragg angle, and $\lambda$ the de Broglie wavelength of the electrons. Often the mosaicity can be quantified by a single angle, $\sigma_\omega$, the standard deviation of rotation angles $\omega$ between the mean orientation and the actual orientations of the mosaic blocks (or "domains"). This assumes that the mosaic spread is isotropic. 

In typical UED experiments, a significant contribution to the mosaic spread is due to flatness imperfections resulting from the mechanical instability of thin plates and difficulties in sample preparation and mounting. Orientation distributions owing to such imperfections are not necessarily isotropic. For example, uniform bending of a flat plate would result in a systematic variation in local orientation (represented by the tangent planes), which is manifestly anisotropic. 

\textsc{Garfield} accounts for the anisotropy of the mosaic spread by assuming that the rotation vectors $\vec{\omega}$, which represent the rotations of the domains from the mean orientation (averaged over the whole sample) to the actual orientation, are distributed according to a three-dimensional normal distribution.
\footnote{A rotation vector $\vec{\omega}$ is an axial vector with length $\omega$ equal to the rotation angle (in radians), and direction parallel to the axis of rotation}$^{,}$\footnote{Differences in domain size and other features of the domains that affect their diffracting power are ignored.}
In general, this orientation distribution, $P_{\vec{\omega}}$, is described by six independent parameters, e.g. the standard deviations $\sigma_i^\mathrm{mos} ~ (i = 1, 2, 3)$ along the principal axes and three Euler angles defining the orientation of the principal axes relative to the laboratory coordinate system (orthogonal axes $X,Y,Z$ with $Z$ parallel to the incident beam).  

The distribution $P_{\vec{h}}$ of the domains' reciprocal lattice vectors $\vec{h}$ induced by the rotations
\footnote{To avoid cluttering of symbols, indices distinguishing rotated RLPs from the mean RLP are suppressed.}
is determined by $P_{\vec{\omega}}$. Viewed as a three-dimensional distribution in reciprocal space, $P_{\vec{h}}$ is degenerate, as it occupies a two-dimensional surface given by a sphere of radius $h = |\vec{h}|$ centered at the origin. Other effects on the intensity profile set aside, diffraction conditions require to consider the intersection of this sphere with the Ewald sphere. Thus, diffraction intensity contributions due to mosaicity will appear as circular arcs. The intensity distribution along the arcs is defined by the conditional distribution obtained by restricting $P_{\vec{h}}$ to the (curved) line of intersection. From this, the expected intensity profiles on the detector can be derived, and their total intensities (integrated along the arcs) as well as the centroid positions can be calculated.   

The exact calculation would be cumbersome and would require numerical integration of functions in three dimensions for all RLPs $\vec{h} = i \, \vec{a}^{\star} + j \, \vec{b}^{\star} + k \, \vec{c}^{\star}$ of interest ($\vec{a}^{\star}$,  $\vec{b}^{\star}$, $\vec{c}^{\star}$ basis vectors of the reciprocal lattice). \textsc{Garfield} exploits the fact that $P_{\vec{\omega}}$ is a normal distribution (by assumption) and introduces an approximation which is only valid if all relevant rotation angles are  sufficiently small. In the following, the approximation is first motivated by considering the case where all rotations are infinitesimally small; subsequently, the question is addressed as to how large the angles of rotation may be so that the approximation can still be used for practical purposes.

Assuming that the rotation vectors $\vec{\omega}$ (defining the orientation of the domains relative to the mean orientation) are normally distributed implies that $P_{\vec{\omega}}$ is a centered trivariate normal distribution of vector coordinates, for example $(\omega_x, \omega_y, \omega_z)$ in the laboratory coordinate system. It is completely defined by its covariance matrix $\Sigma_{\vec{\omega}}$, which can be expressed in terms of the standard values and Euler angles mentioned above. In the non-degenerate case, its three-dimensional density function can be written as:
\footnote{Distributions and their density functions in various parameterizations are denoted by the same symbol. The context and the names of the variables used should remove any ambiguity.}
\begin{equation} \label{eq11}
P_{\vec{\omega}}(\omega_x, \omega_y, \omega_z) = 
\frac{\exp{ \left\{-\frac{1}{2} (\omega_x, \omega_y, \omega_z) \big(\Sigma_{\vec{\omega}}^{(xyz)}\big)^{-1} (\omega_x, \omega_y, \omega_z)^{\top} \right\} }}
{\sqrt{(2 \pi)^3 \, \big|\Sigma_{\vec{\omega}}^{(xyz)}\big|}} 
\end{equation} 

The first step in calculating the expected intensity profile of the reflection corresponding to a given RLP $\vec{h}$ is to transform the density function of $P_{\vec{\omega}}$ from the laboratory coordinate system to a local Cartesian coordinate system  $(\omega_1, \omega_2, \omega_3)$ with the third axis parallel to $\vec{h}$. The other two axes are chosen to be perpendicular ($\omega_1$) and parallel ($\omega_2$) to the scattering plane, which is spanned by $\vec{h}$ and the wavevector $\vec{k_\mathrm{i}} \parallel Z$  of the incident beam ($|\vec{k_\mathrm{i}}| = 1/\lambda$). The covariance matrix of the transformed density function will be denoted by $\Sigma_{\vec{\omega}}^{(123)}$.

If the rotation corresponding to $\vec{\omega} = (\omega_1, \omega_2, \omega_3)$ is infinitesimal, it is equivalent to three  consecutive rotations with angles $\omega_1$, $\omega_2$, $\omega_3$ around the three axes of the coordinate system. Since rotation about the third axis does not move $\vec{h}$, the effect of $\vec{\omega}$ is the same as that of $\vec{\omega}_\perp = (\omega_1, \omega_2, 0)$. \textsc{Garfield} assumes that the same holds, at least approximately, for all rotations of significant probability density. Then, $P_{\vec{\omega}}$ can be replaced by its marginal distribution, $P_{{\vec{\omega}}_\perp}^{{\perp}\vec{h}}$, obtained by integration over $\omega_3$. The result is a centered, bivariate normal distribution, the covariance matrix of which is obtained from $\Sigma_{\vec{\omega}}^{(123)}$ by dropping both the third row and third column:\begin{equation} \label{eq12}
P_{{\vec{\omega}}_\perp}^{{\perp}\vec{h}}(\omega_1, \omega_2) = 
\frac{\exp{ \left\{-\frac{1}{2} (\omega_1, \omega_2) \big(\Sigma_{\vec{\omega}}^{(12)}\big)^{-1} (\omega_1, \omega_2)^{\top} \right\} }}
{\sqrt{(2 \pi)^2 \, \big|\Sigma_{\vec{\omega}}^{(12)}\big|}} 
\end{equation} 

From the marginal distribution $P_{{\vec{\omega}}_\perp}^{{\perp}\vec{h}}$, a density function for the rotated $\vec{h}$ vectors can be derived  straightforwardly. Within the infinitesimal rotations approximation, the distributions $P_{\vec{h}}$ and $P_{{\vec{\omega}}_\perp}^{{\perp}\vec{h}}$ are interchangeable, and the density function in equation \ref{eq12} is also valid for $\vec{h}$ if variables $(\omega_1, \omega_2)$ are interpreted as a special parameterization of $\vec{h}$. 

An infinitesimal rotation $\vec{\omega}$ transforms an arbitrary vector $\vec{h}$ by adding an infinitesimal vector $\vec{\omega} \times \vec{h}$ to $\vec{h}$. Thus, in this approximation, the RLPs are distributed in a plane perpendicular to $\vec{h}$ (\textit{tangent plane approximation}). The coordinates of the rotated vector corresponding to $(\omega_1, \omega_2)$ are given by $(h_1, h_2, h_3) = h (\omega_2, -\omega_1, 1)$, where $h$ is the length of the original $\vec{h}$ at mean orientation ($\omega_1 = \omega_2 = 0)$. Substitution of variables $(\omega_1, \omega_2)$ by new variables $(\rho_1=h \, \omega_2, ~ \rho_2=-h \, \omega_1)$ in equation \ref{eq12} yields: 
\begin{equation} \label{eq13}
P_{\vec{h}}(\rho_1, \rho_2) = 
\frac{\exp{ \left\{-\frac{1}{2} (\rho_1, \rho_2) \, \Sigma_{\vec{h}}^{-1} \, (\rho_1, \rho_2)^{\top} \right\} }}
{\sqrt{(2 \pi)^2 \, \big|\Sigma_{\vec{h}}\big|}} ,
\end{equation}
with covariance matrix  
\begin{align}  \label{eq14}
\Sigma_{\vec{h}} &= h^2 
\left[ {\begin{array}{rr}
     \Sigma_{22} & -\Sigma_{12} \\
    -\Sigma_{12} &  \Sigma_{11} \\
\end{array} } \right]
\end{align}
defined \textit{via} the matrix elements of $\Sigma_{\vec{\omega}}^{(12)} = \left[ {\begin{smallmatrix} \Sigma_{11} & \Sigma_{12} \\  \Sigma_{12} & \Sigma_{22}  \end{smallmatrix} } \right] $.

In the tangent plane approximation, $\rho_1$ and $\rho_2$ can be identified with $h_1$ and $h_2$, the first two coordinates of $\vec{h}$ in the rotated coordinate system.

The expected intensity profile of the reflection corresponding to the RLP $\vec{h}$ can be obtained (up to a scaling factor containing the squared amplitude of the structure factor) by calculating $P_{\vec{h}}(h_1, h_2)$ along the trace of the Ewald sphere in the tangent plane. To a very good approximation, this trace is a straight line. The result is a normal distribution whose mean value (center position) and variance can be calculated from $\Sigma_{\vec{h}}$ by using another transformation of the coordinate system, such that the trace of the Ewald sphere is parallel to one of the new axes.

For non-infinitesimal rotations, the tangent plane approximation breaks down, and the curvature of the support of $P_{\vec{h}}$, i.e. the sphere of radius $h$ around the origin, has to be taken into account. This is done in \textsc{Garfield} by applying \label{geocorr}geometric corrections that should perform well for all practical purposes
\footnote{Rotation angles $\omega >$ 90° need not be taken into account. The limitations of the correction lie in the approximation of the trace of the Ewald sphere in $(\rho_1, \rho_2)$ space by a straight line within the region of significant density $P_{\vec{h}}(\rho_1, \rho_2)$. This approximation is excellent for RPLs $\vec{h}$ tilted by angles less than $\omega_{\mathrm{tilt}} =$ 15° from the $X,Y$ plane. Tilt angles up to $\omega_{\mathrm{tilt}} =$ 30° or 40° should be acceptable, except where high accuracy of estimated intensities is required. (Here, it must be assumed that the angular spread of $P_{\vec{h}}$ is not much larger than $\omega_{\mathrm{tilt}}$.) As the inclination angle increases, the curvature increases, and the quality of the approximation gradually decreases. As the tilt approaches 90°, that means $\vec{h}$ is on the $Z$ axis, the trace shrinks to a circle of radius $\approx h \, \omega_{\mathrm{tilt}} = h \, \pi / 2$ (exact equality holds if the deviation of the Ewald sphere from the $X,Y$ plane is neglected).}.
Hence, the most critical step in estimating diffraction intensities is the \textit{infinitesimal rotations approximation} used for the transition from the original distribution $P_{\vec{\omega}}$  to its marginal distribution $P_{{\vec{\omega}}_\perp}^{{\perp}\vec{h}}$ and the identification of $P_{{\vec{\omega}}_\perp}^{{\perp}\vec{h}}$ with  $P_{\vec{h}}$.

\begin{figure}[ht]
\centering
\includegraphics[width=7cm]{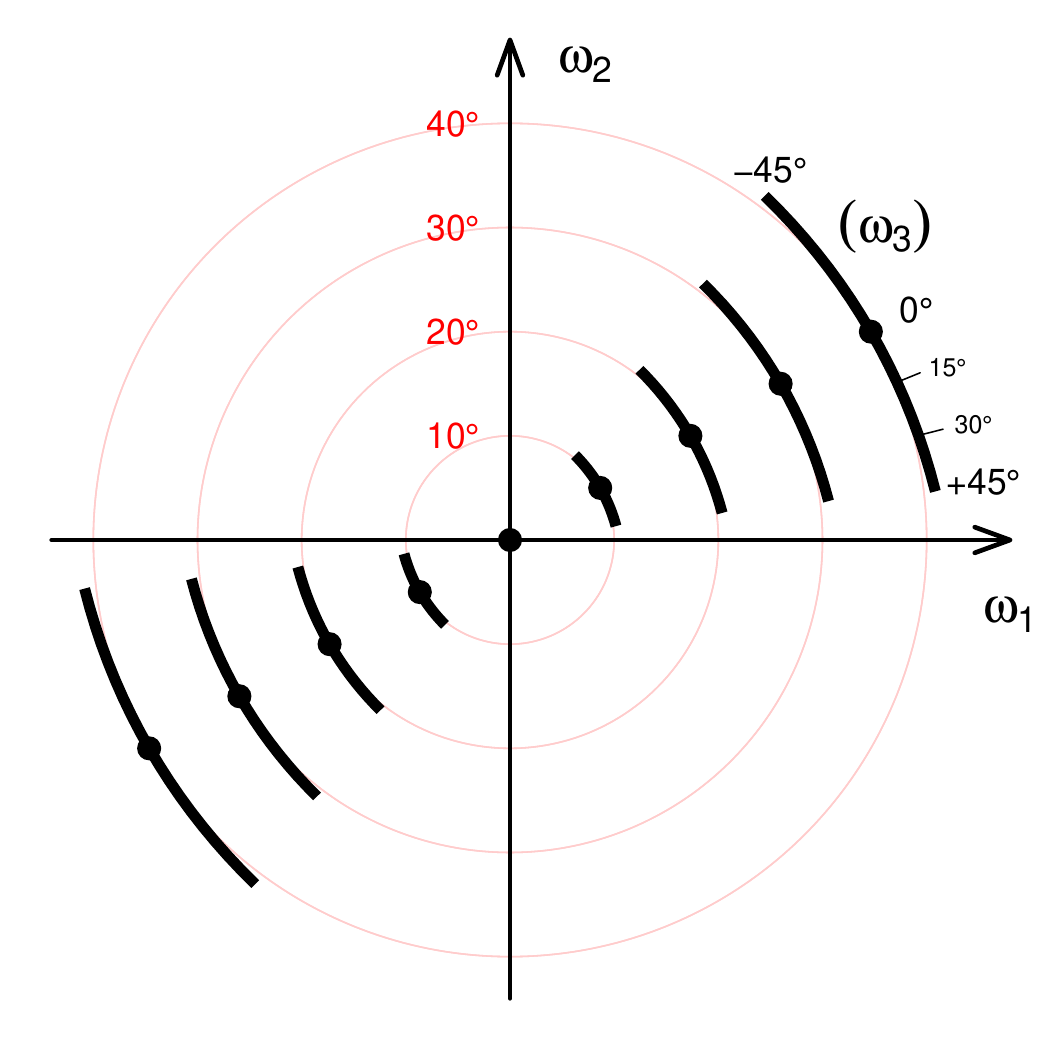}
\caption{Error due to marginalization of $\omega_3$ at non-infinitesimal rotations. A couple of rotations with axes perpendicular to an RLP (with $\vec{h} \parallel \omega_3 ~  \text{axis}$) and rotation angles up to 40° are marked by a series of points in the $(\omega_1, \omega_2)$ plane (black dots). The curved, nearly circular line segments centered at the dots are the projections of all rotation vectors with $\omega_3 \in [-\pi/4, \pi/4]$ rotating $\vec{h}$ into the same direction (see text for further explanations).}
\label{fig:LargeAnglesMix}
\end{figure}

The limitations of this approximation are illustrated in Figure \ref{fig:LargeAnglesMix}. The density of the marginal distribution at any point of the $(\omega_1, \omega_2)$ plane is obtained by integration of $P_{\vec{\omega}}$ along the $\omega_3$ axis, which is parallel to a certain RPL's $\vec{h}$ vector. However, this is not exactly what is required as rotation vectors $(\omega_1, \omega_2, 0)$ and $(\omega_1, \omega_2, \omega_3 \ne 0 )$  do not move the RLP exactly to the same position, unless $\omega_3$ is infinitesimal. In order to integrate densities of rotations that move the RLP to identical positions, $P_{\vec{\omega}}$ should be integrated along curves such that all points of these curves correspont to rotations that result in the same position of the RLP. Projections of such curves to the $(\omega_1, \omega_2)$ plane are shown in Figure \ref{fig:LargeAnglesMix}. 

One observation that follows directly from looking at Figure \ref{fig:LargeAnglesMix} is that the marginal distribution intermixes orientations ($\vec{\omega}$ vectors) within angular ranges that increase almost linearly with $\omega_3$, leading to a certain alteration of the correct distribution. The figure also shows that intermixing of nonequivalent orientations has virtually no effect if the original distribution is isotropic, and that errors due to intermixing increase with the degree of anisotropy. Except in extreme cases of anisotropy, reducing the marginal distribution to a line in the $(\omega_1, \omega_2)$ plane (for example the line through the black dots in Figure \ref{fig:LargeAnglesMix}), standard deviations $\sigma_i^\mathrm{mos}$ up to 5° (i.e. 3 sigma $<$ 15°) should be acceptable, considering that \textsc{Garfield}'s requirements for precision of position and intensity estimates are not very high. Even sigma values in excess of 10° may be tolerable if the anisotropy is moderate.  

\subsubsection{Beam divergence}

In \textsc{Garfield}, the angular spread of the incident wavevectors, $\vec{k}_{\mathrm{i}}$, is modeled as a rotationally symmetric normal distribution with sigma values $\sigma_x = \sigma_y = \sigma^{\mathrm{divg}}$. The effect of a finite beam divergence on the diffraction image is treated approximatively by splitting the effect in two parts and treating them separately. (i) The first effect is that more reflections can appear because more RLPs may "touch" the bundle of Ewald spheres corresponding to the distribution of $\vec{k}_{\mathrm{i}}$ vectors, and the intensities of all reflections change according to the change of "mean" or "effective" excitation errors. (ii) The second part concerns the reflections' positions and profiles in the detector plane, which are affected by beam divergence due to the range of projection directions associated with the angular spread of the $\vec{k}_{\mathrm{i}}$ vectors.

(i) The first effect is taken into account by replacing the distribution $P_{\vec{\omega}}$ of orientation vectors (the "mosaicity") by the convolution of $P_{\vec{\omega}}$ with the distribution describing the divergence of the $\vec{k}_{\mathrm{i}}$ vectors.
\footnote{The convolution of two normal distributions with covariance matrices $\Sigma_1$ and $\Sigma_2$ is a normal distribution with covariance matrix $\Sigma_1 + \Sigma_2$. This is a general result that holds for multivariate and degenerate normal distributions. If the dimensions of $\Sigma_1$ and $\Sigma_2$ do not match, one of them can be filled up with rows and columns of zeros.}
Thus, this part is treated by replacing the mosaicity with an "effective mosaicity", while the Ewald sphere construction remains unchanged.

(ii) In the calculation of the cost function (for parameter fitting), the second effect is ignored. The cost function only depends on the center positions and integrated intensities of predicted reflections. The profiles are irrelevant, and the shift in center positions is negligible if the beam divergence is significantly smaller than the mosaicity, which is the case in typical UED experiments. Nevertheless, for simulating the diffraction \textit{image}, the broadening of reflection profiles by $\sigma^{\mathrm{divg}}$ is approximately taken into account.  

\subsubsection{Energy bandwidth}

An energy distribution around the mean electron energy leads to a distribution of $k_{\mathrm{i}} = 1/\lambda$ values and thus to a distribution of Ewald spheres with radii $k_{\mathrm{i}}$. Again, \textsc{Garfield} uses a normal distribution (sigma value $\sigma^{\mathrm{bwdth}}$) to describe the effect of a finite bandwidth. Similar to beam divergence, the energy bandwidth affects (i) which RLPs give rise to a reflection and with what intensity, and (ii) where in the diffraction image the reflections appear and what their intensity profile is.
\textsc{Garfield} treats the effect of energy distribution in an approximative manner, by separating the two aspects. This leads to another increase of the effective mosaicity by convolution with the distribution describing the energy spread. In this case, however, the modification of $P_{\vec{\omega}}$ depends on the resolution, i.e. the distance of the RLP from the origin, $|\vec{h}|$. For the same reason as in the treatment of beam divergence, the second effect (ii) is considered only when simulating diffraction \textit{images} (where reflection profiles should be as realistic as possible). For parameter fitting and  predicting diffraction \textit{patterns} the second effect is ignored.
\footnote{The fact that some effects of beam divergence and energy bandwidth can be neglected in parameter fitting, but not in simulation of diffraction images, has a corollary: some of the available model parameters can hardly be fitted because the cost function does not strongly depend on these parameters. They are not strictly necessary for indexing with \textsc{Garfield}. However, if realistic diffraction images are desired (once the essential parameters have been established), it may be necessary to guess the values of non-essential parameters and adjust them by trial and error.}

\subsection{Further approximations and computational shortcuts}\label{sec:model.4}

Using normal distributions to describe the four effects discussed in section \ref{sec:model.3} has the advantage that predicting reflection intensities and positions is largely reduced to algebraic manipulations of covariance matrices. The combination of beam divergence and energy spread with the orientation distribution of coherently diffracting domains is not a major complication and can be achieved by convolution, resulting in an effective mosaicity. This is unproblematic because it describes the broadening of the RLPs' diffracting profiles due to three effects that are all of the same kind: all three are related to relaxing one of the conditions of an ideal diffraction experiment with scattering geometry defined by a unique direction of the incident beam, a unique energy of the electrons, and a unique orientation of the crystal lattice.

The finite size effect is of a different kind. Unless the (average or effective) shape function is spherically symmetric, it is not sufficient \textendash~in principle \textendash~to know $P_{\vec{h}}(\vec{h})$, the density of RLPs at any point $\vec{h}$, without also considering the orientation of the corresponding subset of crystallites, which deviates from the average orientation of the crystal lattice by virtue of $\omega_1 = -\rho_2 / h$, $\omega_2 = \rho_1 / h$, and  $\omega_3$. Within the \textit{tangent plane approximation}, rotation of the shape function (represented by normal distribution $P_{\mathrm{shp}}$) due to $\omega_1$ and $\omega_2$ can be neglected. Thus, for this part, a simple convolution of $P_{\vec{h}}$ with $P_{\mathrm{shp}}$ would  suffice. Extending this approach to larger values of $\omega_1$ and $\omega_2$ should be possible by means of the geometric corrections mentioned in the previous section. 
\footnote{Here, it is assumed that the orientation of the effective shape function, $P_{\mathrm{shp}}$ follows that of the crystallites. This seems to be the most natural assumption for samples like thin foils that are curved or corrugated. However, it is difficult to make a general statement about the correlation between orientation and shape of the crystallites that applies to all types of UED samples.}

With regard to $\omega_3$, an exact treatment is computationally more expensive, since all information about $\omega_3$ rotations is lost when calculating $P_{\vec{h}}(\omega_1, \omega_2)$, the marginal distribution of $P_{\vec{\omega}}(\omega_1, \omega_2, \omega_3)$. The exact treatment would involve three steps: instead of using $P_{\vec{h}}(\omega_1, \omega_2)$, consider the conditional distributions $P_{\vec{\omega}}(\omega_1, \omega_2 \,|\, \omega_3)\,\mathrm{d}\omega_3$, calculate their convolution with $P_{\mathrm{shp}}$ rotated by $\omega_3$, and numerically integrate the results over $\omega_3$. We considered such a protocol, but the results did not justify the extra computational costs associated with it, and in the current version of \textsc{Garfield} the rotation of the shape function due to $\omega_3$ is ignored.
\footnote{A clear difference in predicted reflection intensities and positions can only be expected  under certain conditions. First, the shape function must have a marked anisotropy. (This could be the standard case for electron diffraction of samples in thin plate-like geometry.) The orientation spread around $\vec{h}$ (i.e. the range of $\omega_3$ values) must not be too small, and the orientation spread perpendicular to $\vec{h}$ (i.e. the range of values $h\,\omega_1$ and $h\,\omega_2$ must not be too large compared to the shape function. Note that the ranges perpendicular to $\vec{h}$ increase linearly with $h$, while the shape function is the same for all $\vec{h}$.) Furthermore, if the above conditions are favorable for a strong $\omega_3$ effect, the expected expansion of the reflections' intensity profiles will be symmetric or almost symmetric for many (if not all) of the relevant RLPs because the $\omega_3$ range is symmetric around $\omega_3 = 0$. Hence, the effect on centroid positions and integrated intensities will be small.}

Convolution of the bivariate (degenerate trivariate) distribution $P_{\vec{h}}$ with the shape function results in a non-degenerate trivariate distribution. This poses no further problems. The result is a two-dimensional intensity profile with non-zero width in radial direction. The radial width is only determined by the shape function.

~

Anisotropy of the orientation distribution appears to be a common phenomenon in UED, but it is not always a dominant feature. Whenever possible, the general treatment described above should be avoided by using a simpler approach that permits much faster calculations by ignoring mosaic anisotropy. In the \textsc{GeoFit} tool and for simulating and displaying diffraction \textit{patterns}, \textsc{Garfield} offers the option to choose between two methods for predicting position and intensity of reflections: the method presented in the previous sections ("ANISO modeling"), and a simplified version called "standard modeling".
\footnote{Standard modeling is the only method used in the \textsc{GridScan} tool and should always be preferred in \textsc{GeoFit} unless there are clear indications that anisotropy in mosaicity needs to be considered. This is usually not the case at the beginning of a project. By contrast, ANISO modeling is the only method used for simulating diffraction images.}

In standard modeling, the four factors that determine the RLPs' diffracting power profiles are all described by Gaussian functions with standard deviations $\sigma^{\mathrm{shp}}$, $\sigma^{\mathrm{mos}}$, $\sigma^{\mathrm{divg}}$, and $\sigma^{\mathrm{bwdth}}$. Here, the orientation spread measured by $\sigma^{\mathrm{mos}}$ is assumed to be the same in all directions (i.e. isotropic), while the shape profile measured by $\sigma^{\mathrm{shp}}$ is restricted to a certain direction ("relrods"). Standard modeling  assumes that the coherently diffracting domains of the sample can be viewed as thin plates. Their finite extent in lateral directions is ignored.
\footnote{In \textsc{GeoFit}, the user can choose whether the direction of the relrods should always be fixed and parallel to the $Z$ axis, or possibly inclined to the $Z$ axis. If inclined, the tilt is defined by two parameters (polar and azimuthal angle), which can be fitted.}

In order to estimate reflection intensities and positions in standard modeling, first, the effects associated with uncertainties in scattering geometry are combined, resulting in an effective mosaicity with variance $({\sigma^\mathrm{MOS}})^2 = ({\sigma^\mathrm{mos}})^2 + ({\sigma^\mathrm{divg}})^2 + ({a\,h\,\sigma^\mathrm{bwdth}})^2$, where $a$ is a constant and $h = |\vec{h}|$. Then, the effective mosaicity is combined with the shape profile to obtain the width (sigma value, $\sigma^{\mathrm{*}}$) of the effective Gaussian profile of an RLP at $\vec{h}$:
\begin{equation} \label{eq15}
(\sigma^{\mathrm{*}})^2 = (\sigma^{\mathrm{shp}})^2 + (h\,\sigma^{\mathrm{MOS}})^2.
\end{equation} 

\begin{figure}[ht]
\centering
\includegraphics[width=4cm]{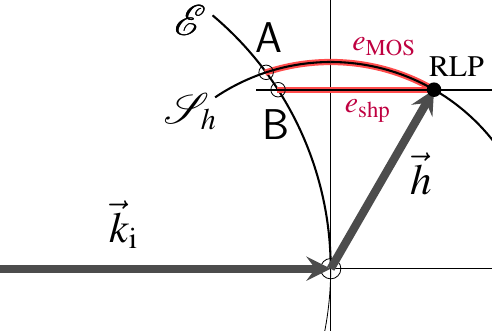}
\caption{Ewald sphere construction to predict the position and intensity of a reflection belonging to an RLP at reciprocal lattice vector $\vec{h}$. 
$\mathcal{E}$ is the trace of the Ewald sphere in the scattering plane which contains the the RLP and the incident wave vector $\vec{k}_{\mathrm{i}}$. Two extreme cases with intersection points $\mathsf{A}$ and $\mathsf{B}$ are shown, as discussed in the main text. The "excitation errors" $e_{\mathrm{MOS}}$ and $e_{\mathrm{shp}}$ are the distances of the RLP from $\mathcal{E}$ measured along the circle around the origin with radius $h$ (i.e. the trace of sphere $\mathcal{S}_h$) and along the direction of the relrods, respectively. Note that the relrod in the figure is parallel to $\vec{k}_{\mathrm{i}}$. This is a special case. In general, relrods can be tilted in any direction. Note also that in electron diffraction, the radius of $\mathcal{E}$ is much larger than that of $\mathcal{S}_h$. }
\label{fig:RLP-to-Ewald}
\end{figure}

Finally, $\sigma^{\mathrm{*}}$ is used to estimate the integrated intensity and centroid position of the reflection associated with the RLP. Since $\sigma^{\mathrm{*}}$ involves two fundamentally different effects, this cannot be done in a straightforward manner. Consider the two extreme cases: (1) $\sigma^{\mathrm{shp}} \ll h \, \sigma^{\mathrm{MOS}}$ and (2) $\sigma^{\mathrm{shp}} \gg h \, \sigma^{\mathrm{MOS}}$, as illustrated in Figure \ref{fig:RLP-to-Ewald}.

(1) If broadening of the RLP profile due to the shape function can be neglected, the profile is determined by the (effective) orientation distribution, so $\vec{h}$ is spread over the surface of the sphere $\mathcal{S}_h$ with radius $h$ around the origin. Without introducing further assumptions, the best guess for the centroid position of the expected reflection corresponds to the point $\mathsf{A}$ on the intersection of the Ewald sphere  with $\mathcal{S}_h$ and the scattering plane spanned by $\vec{k}_{\mathrm{i}}$ and $\vec{h}$. Let $e_{\mathrm{MOS}}$ be the distance from the RLP to $\mathsf{A}$, measured along $\mathcal{S}_h$ in the scattering plane. Then the intensity can be estimated by the value of the profile at $e_{\mathrm{MOS}}$ (up to a factor containing the structure factor amplitude squared).

(2) If the effective mosaicity is negligible compared to the finite size effect and the shape function is that of a thin plate, $\sigma^{\mathrm{*}}$ represents the diffracting power profile of relrods, which are usually more or less parallel to the (mean) surface normal of the crystal foil. The expected centroid position of the reflection is then determined by the point $\mathsf{B}$ where the straight line through the tip of $\vec{h}$ parallel to the direction of the relrods hits the Ewald sphere. The estimated intensity is proportional to the value of the Gaussian density profile at $e_{\mathrm{shp}}$, the distance between $\mathsf{B}$ and the tip of $\vec{h}$. 

In standard modeling, \textsc{Garfield} calculates intensities and centroid positions for both extremes and takes a weighted average (with weights proportional to $\sigma^{\mathrm{shp}} / e_{\mathrm{shp}}$ and $h \, \sigma^{\mathrm{MOS}} / e_{\mathrm{MOS}}$) as estimate for the general case.

\section{Conclusion}\label{sec:conclusion}

\textsc{Garfield} is an interactive software package that helps to (A) find the $hkl$ indices of Bragg reflections recorded in UED experiments and (B) to quantify the intensity contributions of individual reflections when several of them overlap to form a merged "Bragg spot". The latter is a common challenge due to the high mosaicity of typical samples in UED. 
Often the orientation distribution dominates the effect of limited thickness, as is the case in the example of Figure \ref{fig:CmpdA_combi}. 

\begin{figure*}[ht]
\centering
\includegraphics[width=\textwidth]{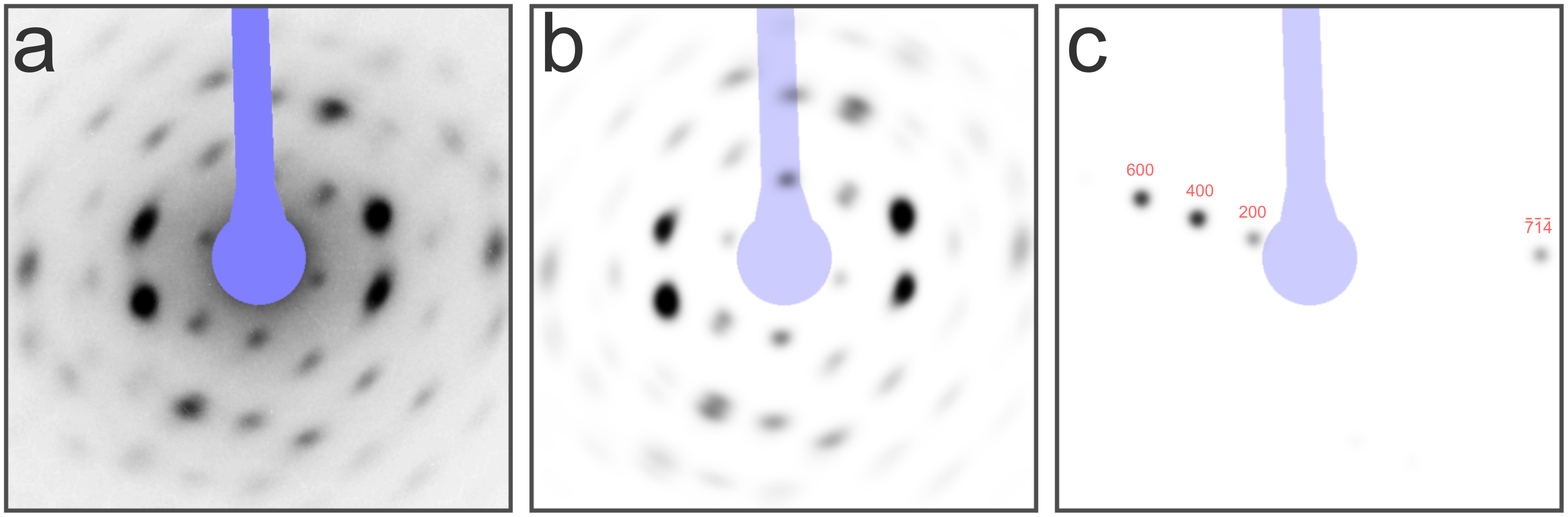}
\caption{Importance of the orientation spread (mosaicity). \textbf{(a)} Recorded and \textbf{(b)} simulated diffraction images as in Fig. \ref{fig:CmpdA_combi}, a and d. 
The mosaicity was fitted with \textsc{GeoFit} to $\sigma_1 = \sigma_2 = 4.7^{\circ}$ ($\sigma_3 = 3^{\circ}$ was fixed, see main text.)
\textbf{(c)} Simulated image computed with the same parameters as in (b), except that the mosaicity was set to zero ($\sigma_1 = \sigma_2 = \sigma_3 = 0$). Note that intensities are always scaled relative to the highest value. The reflections in (b) are also present in (c), but their contribution to the observed diffraction is very weak. For example, the spot near the reflection (400) is actually dominated by the (402) reflection.
}
\label{fig:CmpdA_EffctOfMos}
\end{figure*}

Neglecting the orientation spread would completely change the character of the expected diffraction, as demonstrated in Figure \ref{fig:CmpdA_EffctOfMos}. Of the large number of recorded reflections, only a few would be expected, and -- what is even worse -- the expected reflections are not the intense ones that really dominate the diffraction and largely determine the center position of the observed spots. 

\textsc{Garfield} relies on approximations of the scattering process such as kinematic diffraction theory and  the assumption of normally distributed quantities to mitigate computational costs when predicting reflection intensities from the values of a certain set of model parameters, as described in section \ref{sec:model}. These approximations are supported by physical arguments and have shown resilience in practical use cases.

To A: Like other indexing tools,  \textsc{Garfield} uses the position of observed reflections to determine the orientation of the crystal lattice. The differences are due to the fact that in the cases envisaged for the application of \textsc{Garfield}, the accuracy of the position data cannot be trusted as much as in X-ray diffraction off high-quality single crystals. The uncertainty in reflection positions is taken into account by defining so-called "ambits" around the Bragg spots. The relaxation of positional constraints through the introduction of ambits can lead to difficulties in discriminating between two or more posssible solutions. Fortunately, such ambiguities can often be resolved by also considering the predicted intensities, even if these intensities have large margins of error.  As an additional quality check of the predictions, visual comparison of simulated and recorded diffraction patterns introduces an element of "cross-validation" that turns out to be crucial. If by all criteria, such as SSR values and R factors, plausibility check of the model parameters, and visual comparison of diffraction images, it is not possible to discriminate between non-equivalent solutions, the task must be considered unsolved.
\footnote{The \textsc{Garfield} software is provided with no guarantee as to the accuracy or the correctness of the results.}

Even if in a particular physical situation the Bragg reflections do not overlap \textendash~such that  single triplets of Laue indices $hkl$ can be assigned to each Bragg spot \textendash~ finding the correct indexing for a single diffraction image can still be a challenge (for any tool) if nonequivalent lattice planes have identical or similar metrics by pseudo-symmetry or by accident. Working interactively with \textsc{Garfield}, especially when using \textsc{GridScan}, increases the chance of becoming aware of potential pitfalls. Using reflection intensities as additional discriminating information will in most cases favor the correct solution even if the intensity estimates are not very accurate.

To B: Superposition of two or more Bragg reflections in a single diffraction peak is not an uncommon situation in UED experiments. If such maxima are to be used to study ultra-fast structural changes by pump-probe experiments, knowledge of the fractional contributions of individual reflections is crucial for the interpretation of measured intensity variations. In the case of overlapping reflections, \textsc{Garfield} provides not only reflection indices for all reflections involved, but also returns estimates of the individual intensity contributions to enable further analysis. 

The validity of the approximations used to estimate individual reflection intensities is crucial if the relative contributions are to be used in the structural dynamics analysis. The same standards should be applied here as for structure analysis with electron crystallography in general. In particular, the sample must be thin enough so that effects of dynamical and multiple scattering are sufficiently suppressed. If this is not the case, \textsc{Garfield}'s estimates of diffraction intensities are compromised. If dynamical scattering can be ignored, but the errors due to other approximations (e.g. the assumption of normal distributions to describe the orientation spread and diffracting power of RLPs) are inadequate, one option might be to find  better estimates, for example by using a better model or by applying more precise calculations. Another possibility could be to separate the contributions of individual Bragg reflections through a detailed analysis of the spot profiles. \textsc{Garfield}'s intensity estimates might be useful as start values for profile fitting.

As input, \textsc{Garfield} needs a list of measured diffraction spot intensities and corresponding coordinates. Information contained in spot profiles is not taken into account. The first step in the analysis of UED images, the extraction of intensities and positions from recorded images, must be done outside of \textsc{Garfield} and prior to using the program. So far, \textsc{Garfield} has little flexibility to ensure an optimal match between the way data were extracted from the recorded diffraction images and \textsc{Garfield}'s particular way of using those data in parameter fitting and indexing: a subset of the input data can be selected by enabling or disabling a mask or by restricting the resolution range to use, and the default ambit radius can be changed if deemed advisable. 
For future releases of \textsc{Garfield}, data reduction tools (including background subtraction, masking of regions, definition and localization of Bragg spots, integration of spot intensities) and the tools for model fitting and indexing provided by the present version should both be made accessible via the same interface for enhanced user experience.

\subsection*{Supplementary Material}
Figures S1 and S2 are presented in the first section of the pdf available as supplementary material. The second part of the pdf contains three illustrative examples for the application of \textsc{Garfield}.

\begin{acknowledgments}
This work was supported by the Max Planck Society. We thank R. J. Dwayne Miller (UToronto) for initiating this work, Stuart A. Hayes (European XFEL), Gast\'{o}n Corthey (UNSAM), and R. Patrick Xian (UToronto) for providing diffraction data on $\mathrm{Me_{4}P[Pt(dmit)_{2}]_{2}}$ and $\mathrm{TBAI_3}$ and for many fruitful discussions, we thank Heinrich Schwoerer (MPSD) and Gabriele Tauscher (MPSD) for providing the rubrene and $\kappa$-(BEDT-TTF)\textsubscript{2}Cu[N(CN)\textsubscript{2}]Br ("alpha-$\kappa$") diffraction data. We are thankful to Kazushi Konada (Universität Stuttgart) and Kazuya Miyagawa (University of Tokyo) for kindly providing the alpha-$\kappa$ crystals as well as to Jens Pflaum and Sebastian Hammer (Universität Würzburg) for allowing us to use their rubrene samples for analysis (Supplementary Material).

\end{acknowledgments}

\section*{Author declarations}

\subsection*{Conflict of Interest}

The authors have no conflicts to disclose.

\section*{Data Availability}
\textsc{Garfield} is available as a collection of \textsf{R} scripts and related files by downloading a zip file (https://doi.org/10.17617/3.CXELBR) from \textsf{Edmond}, the Open Research Data Repository of the Max Planck Society, and is provided under the GNU GPLv3 licence terms. 

\bibliography{GFrefs_02}

\end{document}


\maketitle

\tableofcontents

\newpage

\section{Working with \textsc{Garfield}: supplemental information}

\subsection{Figure S1: The main window}

\begin{figure*}[ht]
\centering
\includegraphics[width=\textwidth]{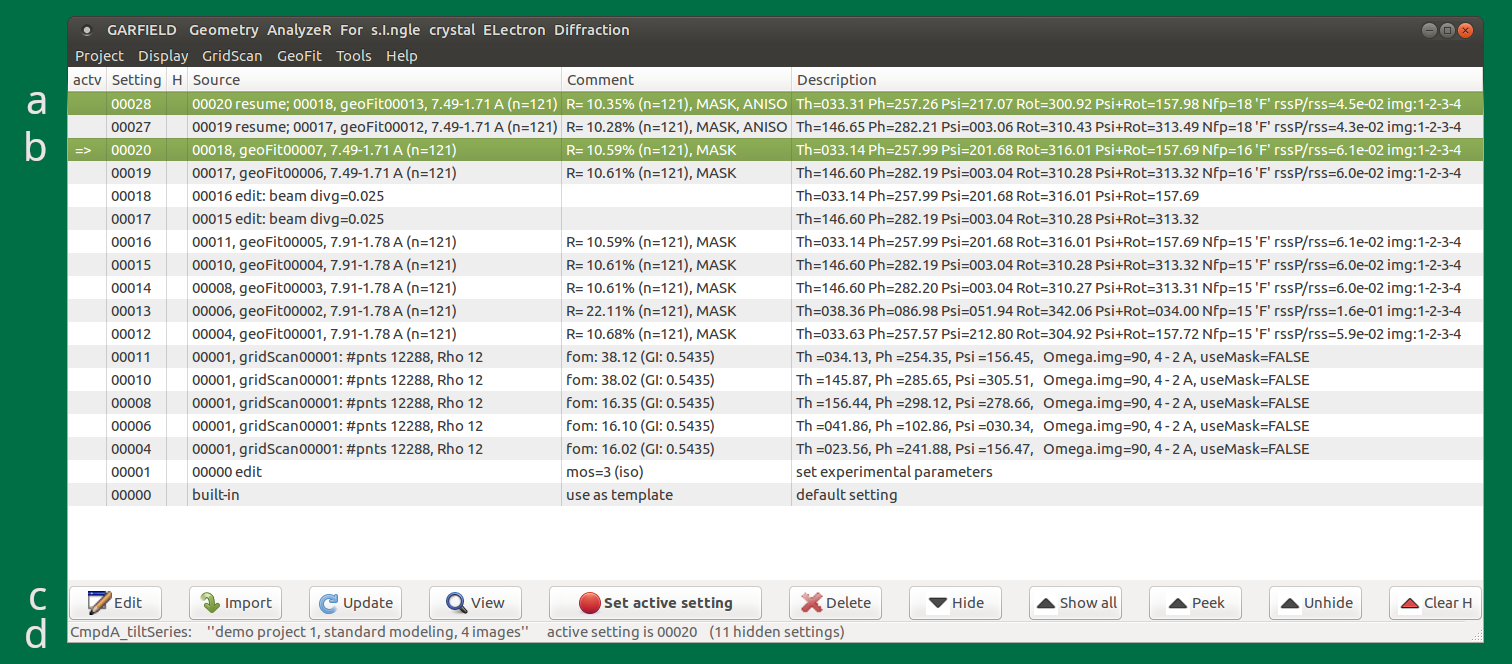}
\label{fig:CmpdA_mainTable}
\end{figure*}
\vspace{-0.7cm}
\begin{center}
\begin{minipage}{15cm}
\onehalfspacing
\small{
\noindent
The main part of the window is a table with columns "actv" (indicating the currently active setting), "Setting" (a unique 5-digit integer ID), "H" (used to label settings that are marked for hiding), and three columns "Source", "Comment", and "Description" that can be edited and filled with arbitrary text to help recognize and identify settings; when a row is automatically created with one of the various tools accessible via drop-down menus, these fields are pre-filled with useful information. Each row of the table corresponds to a complete set of model and control parameters required for simulating diffraction. If one or several rows are highlighted \textbf{(a, b)} these settings are selected for actions started with the "action buttons" \textbf{(c)} below the table. The line with the "{\tt =>}" symbol in the first column \textbf{(b)} represents the "active setting", the set of parameters that are currently used for calculation and simulation of images and patterns. The Edit button enables viewing and editing individual parameters of a setting, as well as creating new settings. The View button displays parameter sets in tabular form for easy comparison of multiple settings. Clicking the "Set active setting" button makes the highlighted setting the "active setting". The status bar \textbf{(d)} shows the name of the current project, a short description entered when the project was defined, the unique ID of the active setting, and the number of hidden settings.}
\end{minipage}
\end{center}

\vspace{0.5cm}

\noindent
When \textsc{Garfield} is started, the main window opens with no content. Once a \textit{project} has been created or selected from a list of previously defined projects, the main window displays a list of "settings" (complete sets of model and control parameters) defined for the current project. Figure S1 shows the main window with a project named "CmpdA\_tiltSeries" opened.
All settings and related information are automatically stored in the file system and remain accessible until the user explicitly deletes settings from the main table.

\newpage
\subsection{Figure S2: The \textsc{GeoFit} User Interface}

\begin{figure*}[ht]
\centering
\includegraphics[width=\textwidth]{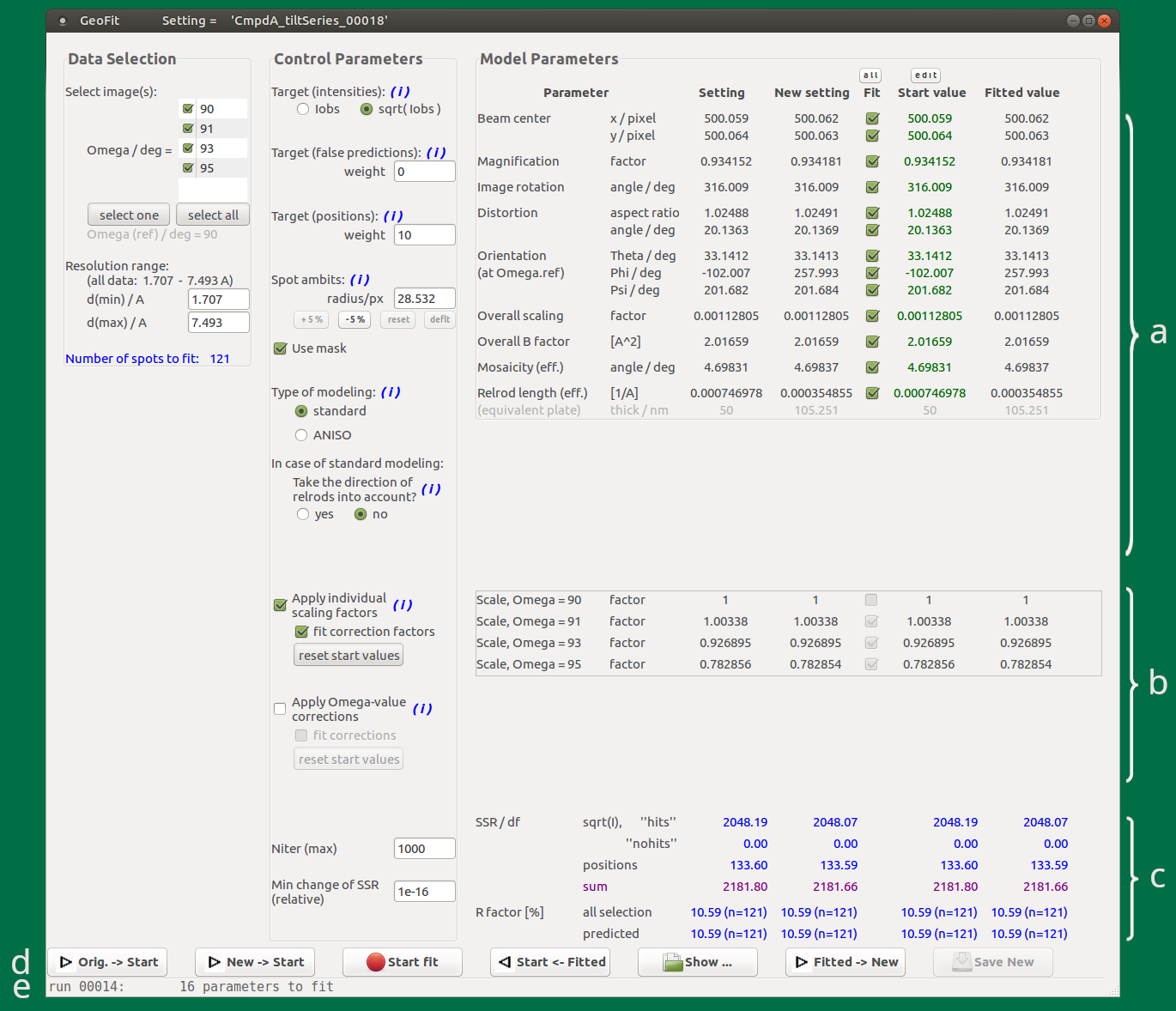}
\label{fig:CmpdA_geofit}
\end{figure*}
\vspace{-0.7cm}
\begin{center}
\begin{minipage}{15cm}
\onehalfspacing
\small{
\noindent
When the \textsc{GeoFit} tool is started, the GUI is filled with the parameters of the \textit{active setting}. Parameters and flags are listed under the headings Data Selection, Control Parameters, and Model Parameters. Model parameters are listed with four values in columns Setting (original), New setting (to be saved), Start value, and Fitted value, all of which are identical at the beginning. Checkboxes next to the start values allow selection of individual parameters for fitting. 
\textbf{(a)}: block of model parameters common to all projects; their number depends on the type of modeling (standard or ANISO). 
\textbf{(b)}: up to two blocks appearing in multi-image projects (tilt-series) when more than one image is selected in the Data selection box; allowing individual intensity scales and rotation corrections to be applied and fitted. 
\textbf{(c)}: crystallographic R factors and SSR values corresponding to the model parameters in (a) and (b). 
\textbf{(d)}: bar of action buttons used to start/stop runs of NLS fitting, copy values of model parameters from one column to another, and save the values of New setting.
\textbf{(e)}: status bar showing the \textsc{GeoFit} run ID of the next (or current) NLS fitting; when a fit run has been started, displaying the progress of the fit.}
\end{minipage}
\end{center}

\vspace{0.5cm}

\noindent
The \textsc{GeoFit} GUI for optimizing model parameters by NLS fitting can be launched from the menu of the same name in the main window. The \textsc{GeoFit} tool runs as a separate process independent of the main \textsc{Garfield} program. By default, the parameter values of the active setting are used as initial values for NLS fitting with the Levenberg-Marquardt algorithm as implemented in the R package minpack.lm \cite{Elzhov2016}. The default start values can be changed by editing the list of parameters in the \textsc{GeoFit} window (see Figure S2). In a typical single-image project, up to 22 model parameters can be selected for simultaneous fitting, by minimizing the cost function described in the main text. 

The progress of the fit is monitored by a pop-up window showing the decrease in the sum of squared residuals (SSR). Once the fit is completed, the success can be assessed by comparing the SSR values and the conventional crystallographic R-factors before and after the fit. If there is little or no improvement and the result is still unsatisfactory, another fit run can be started with different start parameters and/or control parameters.

\newpage
\section{Additional samples evaluated with \textsc{Garfield}}

\subsection{$\kappa$-(BEDT-TTF)\textsubscript{2}Cu[N(CN)\textsubscript{2}]Br}

\subsubsection{Data description and results}

Electron diffraction of the organic superconductor $\kappa$-(BEDT-TTF)\textsubscript{2}Cu[N(CN)\textsubscript{2}]Br (Kini et al., 1990, \cite{Kini1990}) was measured at room temperature with 48 keV electrons, using a crystal slice of approximately 200 x 400 micrometers with a nominal thickness of 30 nanometers. The crystal structure at 298 K was determined by Kini et al. (1990) \cite{Kini1990} and confirmed by Hiramatsu et al. (2015) \cite{Hiramatsu2015}.

\textit{Data reduction:} The positions and intensities of 71 reflections ("observed diffraction spots") were extracted from the recorded diffraction image by summation of electron counts in elliptical regions around the observed spots after subtracting a radially symmetric background. Two of the spots close to the beam stop were masked and discarded from the calculations.

\textit{Crystal structure:} Crystals of the compound at 298 K are orthorhombic with space group Pnma. Structure parameters by Hiramatsu et al. (2015) \cite{Hiramatsu2015} were used to determine the orientation of the crystal slice with \textsc{Garfield} (Cambridge Structural Database, CSD code JESDUG09, CCDC 1016184): $a = 12.9653$~\r{A}, $b = 30.028$~\r{A}, $c = 8.5444$~\r{A}.

\textit{Results:} Of the 69 accepted Bragg spots, 66 were predicted with a non-zero intensity, using 11 parameters in standard modeling and an ambit radius of 15 pixels. The remaining three spots are all of very low intensity and have potentially been missed because the errors in measured and predicted positions exceed the ambit radius. The fitted orientation parameters are: $\Theta = 54.5$°, $\Phi = 66.4$°, and $\Psi = 49.5$°, where $\Theta$ and $\Phi$ are the polar and azimuthal angles of the incident electron beam relative to the crystal coordinate system (polar axis parallel to the $\mathbf{c}^*$, azimuthal angle measured from $\mathbf{a}$). The incident beam is parallel to the vector [uvw] = [0.3705, 0.3662, 1] (coefficients of the basis vectors, $\mathbf{a}$, $\mathbf{b}$, $\mathbf{c}$) or (hkl) = (0.1886, 1, 0.2211) (coefficients of the reciprocal basis vectors, $\mathbf{a}^*$, $\mathbf{b}^*$, $\mathbf{c}^*$). The mosaicity was fitted to 2.4° (SD) for angular spread around X and Y (perpendicular to the incident beam); the angular spread around the incident beam was fixed and set to 1.5° based on visual inspection of the diffraction data. The fitted thickness was 36 nanometers and represents a mean-like value of the thickness distribution of the different crystallites contributing to the diffraction data. The conventional R-factor (based on square roots of the 66 observed and predicted intensities) was 14.9\%.

\newpage

\subsubsection{Figure Smp1}
\begin{center}
\includegraphics[width=5.4in]{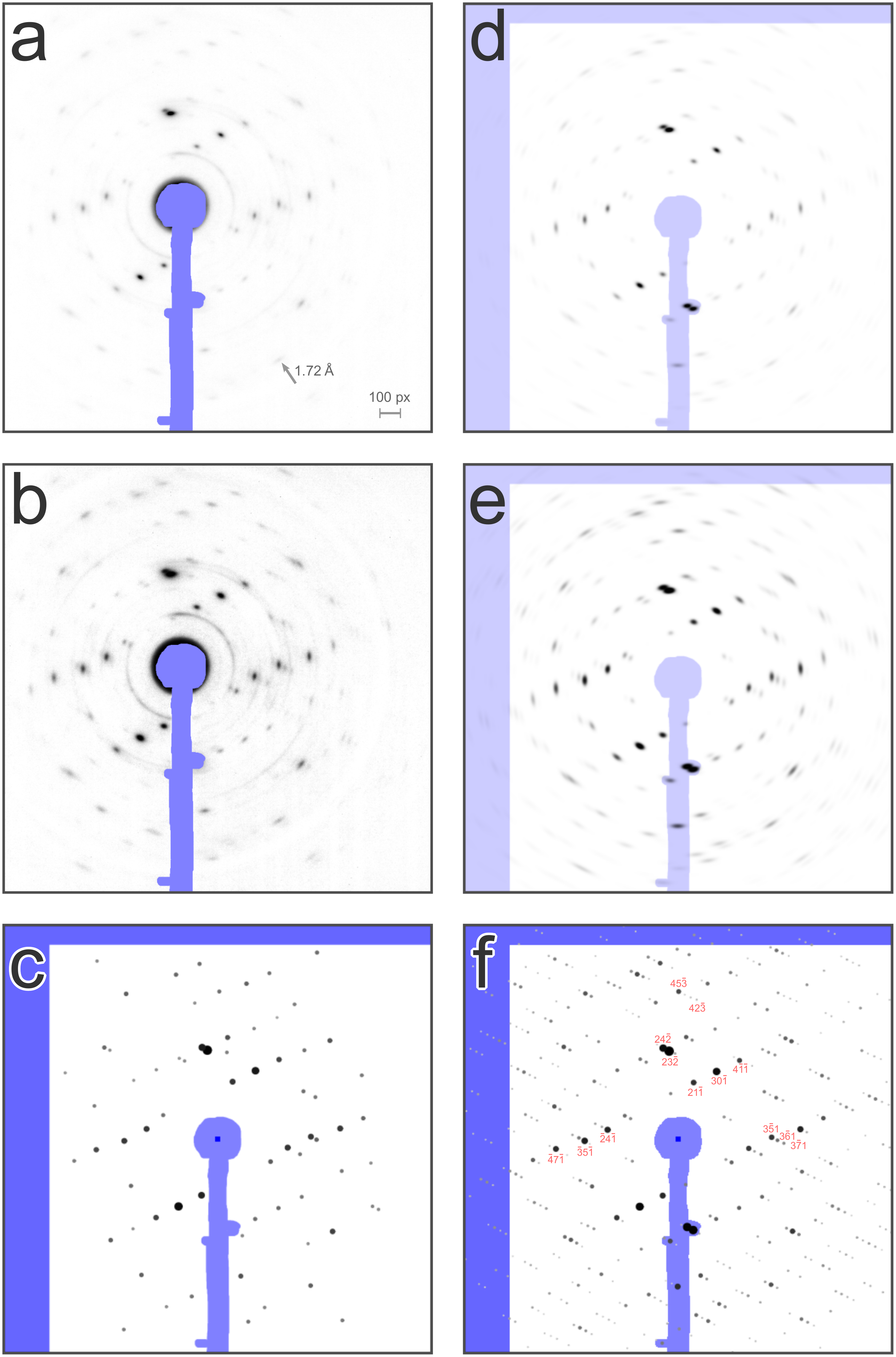}
\end{center}
\newpage


\subsubsection{Legend to Figure Smp1}
\vspace{0.25in}

Observed (a, b, c) and simulated diffraction (d, e, f) of $\kappa$-(BEDT-TTF)\textsubscript{2}Cu[N(CN)\textsubscript{2}]Br; the diffraction images shown in (a, b) are background corrected by subtracting a smooth, radially symmetric function; (a, d): linear intensity scaling; (b, e) with contrast enhancement, gamma = 3.31; (c, f): diffraction patterns corresponding to the observed and simulated diffraction images.  \quad All panels show screenshots of graphics windows displayed by various tools of \textsc{Garfield}. In panel (f) Laue indices have been added for some representative spots. Note that the measured diffraction in (a, b, c) has not been centered, deliberately, in order to include weak diffraction at high-resolution on the right side. The simulated diffraction (d, e, f) is always displayed with the incident beam at the center of the graphics window. As a consequence, part of the simulated area exceeds the observed range; this area is automatically masked by \textsc{Garfield} (blue stripes at the top and the left border of d, e, f) and never used in fitting. Also displayed, with blue shade, is the mask used to exclude the beamstop and few spots that were included in the list of reduced data, although the intensities are not very reliable because the spots are close to the beamstop.

\newpage

\subsection{Tetra-\textit{n}-butylammonium triiodide ($\mathrm{TBAI_3}$)}

\subsubsection{Data description}

The data used to prepare Figures Smp2A, Smp2B, and Smp2C in this section are taken from a short tilt series (five images covering the range from -4° to 4° relative to the reference orientation) that has been recorded to determine the crystal orientation in a study of the bi-molecular reaction dynamics of triiodide formation in crystals of tetra-\textit{n}-butylammonium triiodide \cite{Xian2023}. Three diffraction images recorded at 0° (perpendicular to the incident electron beam), 2°, and 4° were selected for the preparation of the figures. The orientation analysis was repeated with \textsc{Garfield} and fully confirms the orientation and indexing found in the original analysis with a precursor of \textsc{Garfield}. 

The orientation analysis was performed using the crystal structure for $\mathrm{TBAI_3}$ at 173 K (temperature of data collection) deposited at the Cambridge Crystallographic Data Centre (W. S. Brotherton, R. J. Clark, Lei Zhu CCDC 870718: Experimental Crystal Structure Determination, 2013, DOI: 10.5517/ccy71pw). The crystal's space group is P$\overline{1}$ with unit cell parameters a = 9.516 Å, b = 15.655 Å, c = 15.886 Å, $\alpha$ = 83.461°, $\beta$ = 74.229°, $\gamma$ = 78.522°.

\newpage

\subsubsection{Figure Smp2A}
\begin{center}
\includegraphics[width=5.4in]{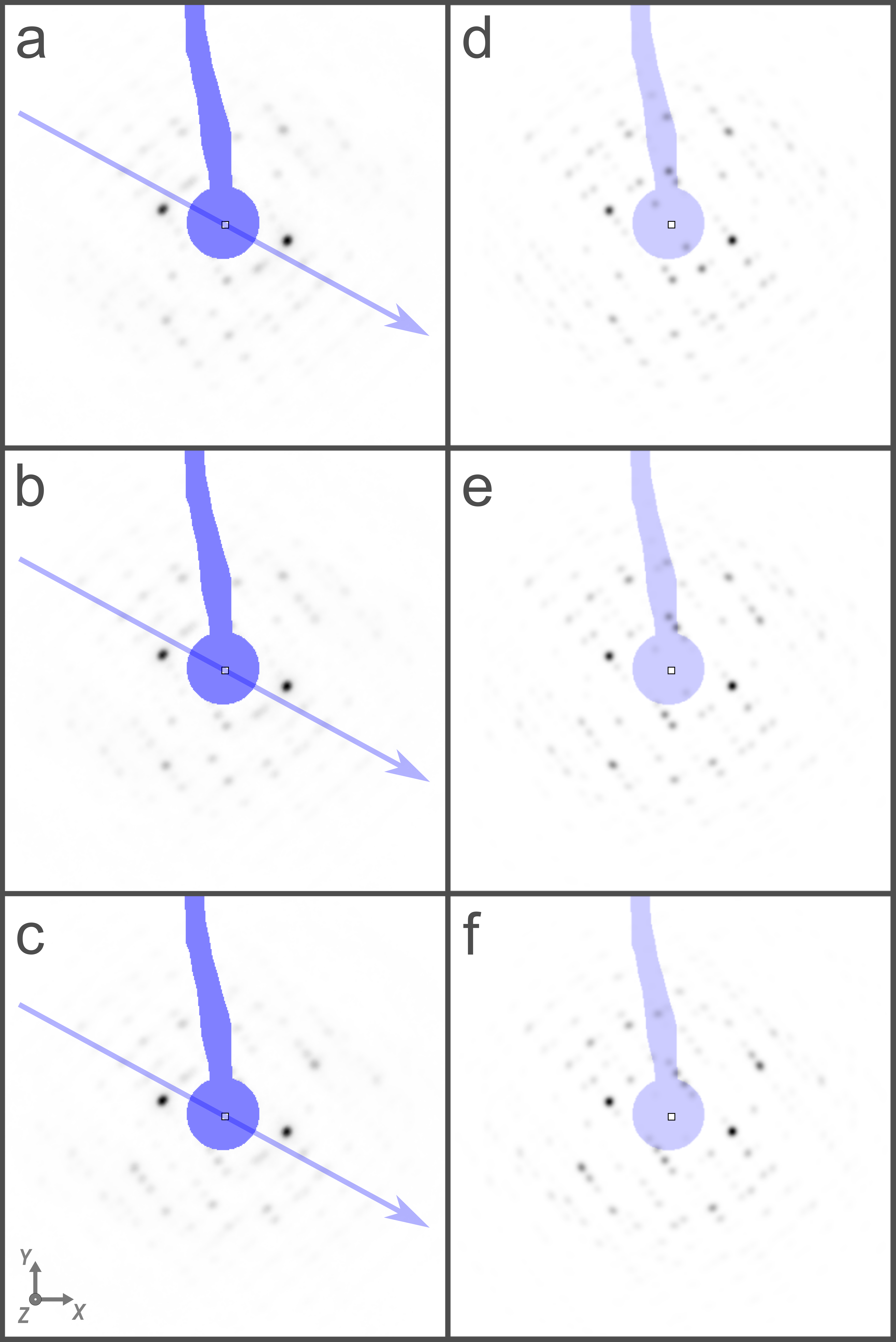}
\end{center}
\newpage

\subsubsection{Figure Smp2B}
\begin{center}
\includegraphics[width=5.4in]{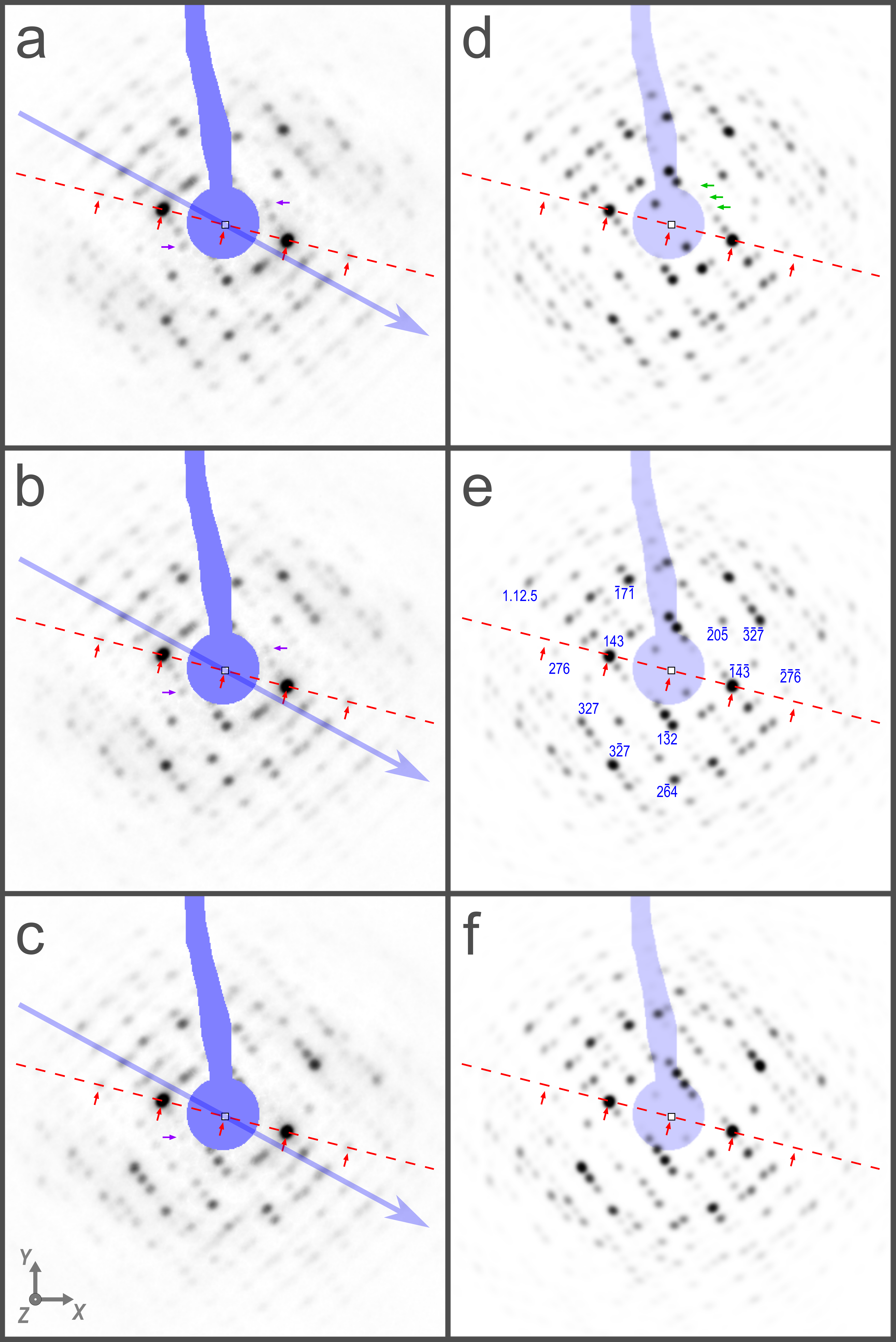}
\end{center}
\newpage

\subsubsection{Legends to Figures Smp2A and Smp2B}
\vspace{0.5in}
\textbf{Figure Smp2A}:
\\ 
\hspace*{0.25in}\textbf{(a, b, c)} Diffraction images of a thin $\mathrm{TBAI_3}$ crystal layer perpendicular to $Z$, the direction of the incident electron beam (\textbf{a}), and after rotation by 2° (\textbf{b}) and 4° (\textbf{c}). The axes of the laboratory coordinate system used by \textsc{Garfield} are shown in panel \textbf{c}. The rotation axis was parallel to $-Y$. Due to the effect of a magnetic lens, the images are rotated around $Z$, such that the apparent direction of the rotation axis (indicated by blue arrows) is at 61° to $-Y$ ($\pm$ 3°). Also shown in blue color is the mask used by \textsc{GeoFit} to exclude the detector region covered by the beamstop.\\
\hspace*{0.25in}\textbf{(d, e, f)} Simulated diffraction images corresponding to the recorded images on the left side and calculated with parameters fitted with \textsc{GeoFit} in a simultaneous fit using spot intensities and center positions from all three recorded images, simultaneously.\\
\hspace*{0.25in}The images in all panels are snapshots taken from \textsc{Garfield}'s display windows using linear intensity scaling (white: zero intensity; black: maximal intensity). The captured images have been corrected by radial background subtraction. Note that the diffraction is dominated by two strong reflections near the rotation axis.\\
\\
\textbf{Figure Smp2B}:\\ 
\hspace*{0.25in}As Figure Smp2A, except that annotations have been added and nonlinear intensity scaling has been applied in order to improve the visibility of weak reflections. The intensity scaling is the same for recorded and simulated images (cut-off at 60\% of the maximum, followed by gamma correction with exponent 3.31).\\
\hspace*{0.25in}In panel \textbf{e}, some of the more prominent reflections are labeled with their Laue indices \textit{hkl}. Note that all visible spots can be assigned a unique Bragg reflection; intensity contributions from other reflections are negligible. The positions of the spots hardly change as the crystal rotates, only the spot intensities change significantly.\\
\hspace*{0.25in}Although the overall agreement between simulation and measurement appears to be more than satisfactory, notable deviations can also be identified. Some of them are marked by small arrows:

\begin{itemize}
  \item \textbf{red arrows} along the central line through the $(1~4~3)$ reflection (all panels)\\
  Spots of weak intensity are observed in the recorded images at positions of the $(2~8~6)$ and $(\overline{2} ~ \overline{8} ~ \overline{6})$ reflections, the second order reflections of $(1~4~3)$ and $(\overline{1} ~ \overline{4} ~ \overline{3})$. However, the calculated intensities of the second order reflections are even much smaller, so they are almost invisible in the simulated images (see also Figure Smp2C, with greater contrast), which is consistent with the fact that the structure factor amplitude of the $(2~8~6)$ reflection is 20 times smaller than that of the $(1~4~3)$ reflection; thus, the intensity ratio expected in kinematic approximation should be on the order of 400.   
  A simple explanation for the relatively high intensity of the apparent second order reflections would be that they are the result of double diffraction of the strong $(1~4~3)$ reflections. Given the weakness of the apparent second-order of the strong $(1~4~3)$ reflection, it can be concluded that double diffraction is probably not the main reason for differences between observed and simulated spot intensities.
  
  \item \textbf{purple arrows} in recorded images (panels \textbf{a}, \textbf{b}, \textbf{c})\\
  These arrows point to a reflection (and its Friedel mate) near the beam center that has no real correspondence in the simulated images. In contrast to the apparent $(2~8~6)$ reflection there is no obvious candidate for double diffraction. The most likely explanation is that this reflection is caused by a parasitic crystal domain.
  
  \item \textbf{green arrows} in the simulated image of panel \textbf{d}\\
  These arrows indicate the continuation of the $(\overline{1} ~ n ~ \overline{3}$) series of reflections (pointing to $n = -1, 0, 1$), the most prominent of which is the $(\overline{1} ~ \overline{4} ~ \overline{3})$ reflection. There is no significant intensity in the recorded images that corresponds to the predicted reflections. This difference can be seen more clearly in the high contrast images of Figure Smp2C. The prediction of non-zero intensities for reflections with large excitation errors is not surprising considering that \textsc{Garfield} uses normal distributions (which extent to infinity) to describe the effect of orientation spread and finite size broadening. Due to the low intensity of these false predictions, their impact on the simulation is negligible.  
  
\end{itemize}

\newpage

\subsubsection{Figure Smp2C}
\begin{center}
\includegraphics[width=5.4in]{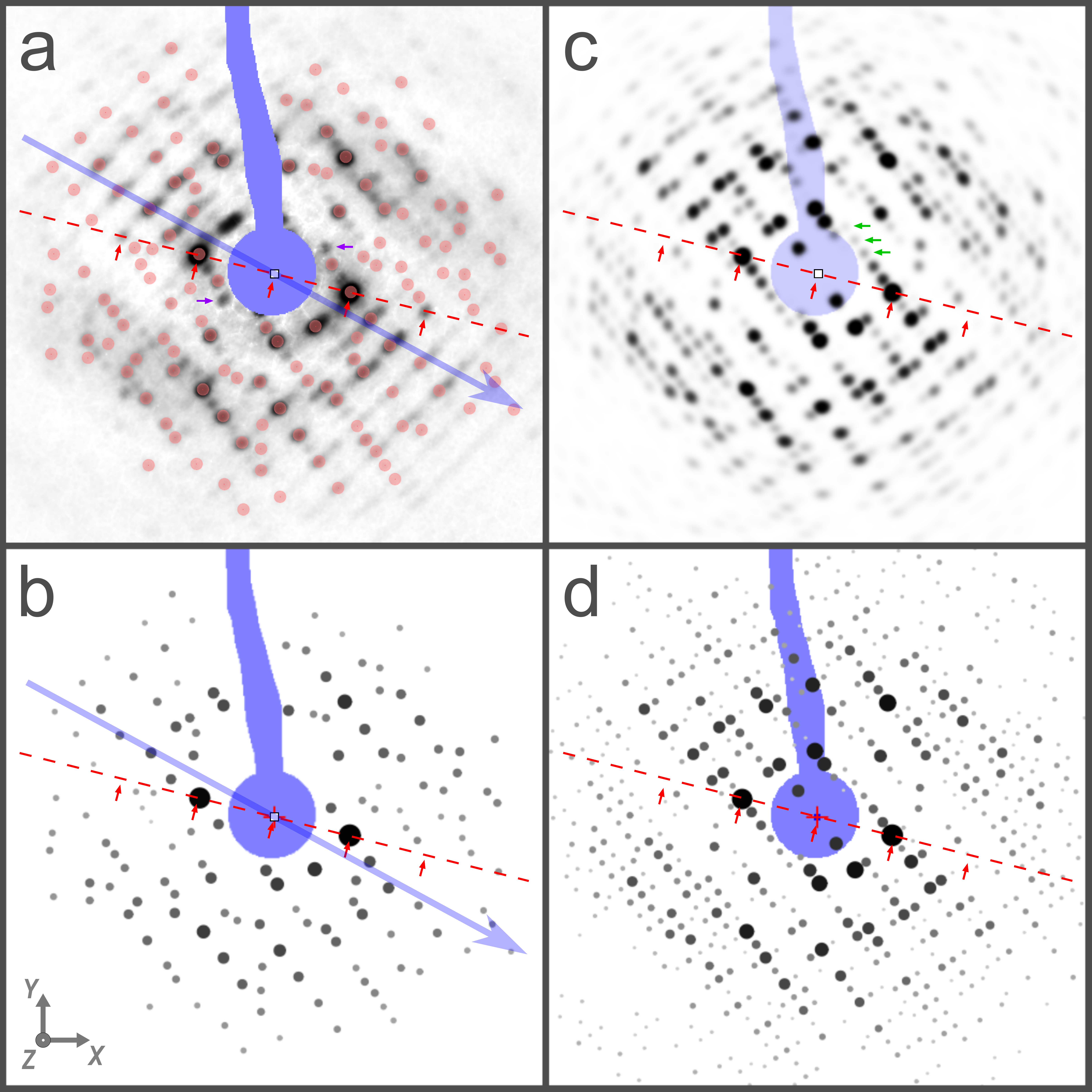}
\end{center}
\textbf{Diffraction images \textit{versus} diffraction patterns at the reference orientation (tilt angle 0°)}\\
\hspace*{0.25in}\textbf{(a, c)} Recorded and simulated diffraction images (with annotations) as in Figure Smp2B (a and d), but with increased contrast (cutoff at 30\% of the maximum intensity, followed by gamma correction with exponent 6.03) to show the differences more clearly. In addition, in panel \textbf{a}, the \textit{ambits} of the reflections that were integrated and included in the list of \textit{reduced data} for use with \textsc{Garfield} are shown as pink disks. \\
\hspace*{0.25in}\textbf{(b, d)} Experimental and simulated diffraction patterns, i.e. graphical representations of the \textit{reduced data} (positions and integrated intensities, $I$, of observed reflections) and the predicted reflections. Reflections are represented by dots of radii and gray values proportional to $I^{1/3}$.\\
\hspace*{0.25in}Each dot in panel \textbf{b} has a corresponding ambit disk in panel \textbf{a}. Note that, for various reasons, not all possible observations have been included in the list of reduced data, most of the time because integration and location of weak reflections is often not reliable. Prediction of observed, but unaccounted reflections confirms the correctness of the solution. The two reflections marked by purple arrows in panel \textbf{a} and attributed to a parasitic crystallite (see legend to Figure Smp2B) were excluded in the final calculations without resulting in a significant improvement. However, it goes without saying that the presence of parasitic diffraction, which is not obvious at the beginning of the analysis, can significantly complicate the first steps towards finding the correct solution or even prevent the correct solution from being found.\\
\hspace*{0.25in}In comparing the predicted pattern in panel \textbf{d} with the simulated image in panel \textbf{b}, one should note that many of the small, low-intensity predictions in \textbf{d}, particularly those at higher resolutions, correspond in the simulation to diffuse, barely visible Bragg reflections.

\vspace{0.5in}
\hspace*{0.25in}The results shown here and in Figures Smp2A, Smp2B were obtained with "standard modeling" (assuming isotropic mosaicity) and without applying any corrections to the tilt angles. A mosaicity of about 2° (fitted value: 1.793°, standard deviation) at a tilt-range of 4° explains why most reflections appear in all three diffraction images at the approximately same positions, albeit with varying intensities. Using ANISO modeling and applying rotation angle corrections did not lead to a significant improvement of the fit, nor did the indexing change.

\newpage

\subsection{Rubrene}

\subsubsection{Data description and background information}

Rubrene (5,6,11,12-tetraphenyl-tetracene, $\mathrm{C_{42}H_{28}}$) crystallizes in at least three different cystal forms: crystals of orthorhombic, triclinic, and monoclinic symmetry have been described in the literature \cite{Henn1971, Jurchescu2006, Huang2010, Matsukawa2010, Fielitz2016, Socci2017, Moret2022}. Thin, rod-shaped rubrene crystals of the predominant orthorhombic form grown by vapor diffusion were kindly provided by Jens Pflaum and Sebastian Hammer (Universität Würzburg). 

Electron diffraction images (35 keV) of two microtome slices (about 50 nanometers thick) cut from these crystals are shown in Figure Smp3A (a, c). 
The data were analyzed with \textsc{Garfield} assuming the orthorhombic crystal structure at 293 K determined by Jurchescu et al. (2006) \cite{Jurchescu2006}, space group symmetry Cmce, point group mmm, a = 26.860\textup{~\AA}, b = 7.193\textup{~\AA}, c = 14.433\textup{~\AA}. 

Expected zone-axis patterns for slices cut along the {001} and the {100} faces are shown in Figure Smp3A (b, d). They capture characteristic features of the recorded images, suggesting that the presumed orientations of the crystal slices are essentially correct. The main difference is that more diffraction spots were observed than are present in the zone-axis patterns, indicating a large orientation spread (apparent mosaicity) of coherently diffracting domains in the slices.

\subsubsection{Figure Smp3A: \quad Diffraction of two slices cut from rubrene crystals}
\vspace{0.5in}

\begin{center}
\includegraphics[width=5.0in]{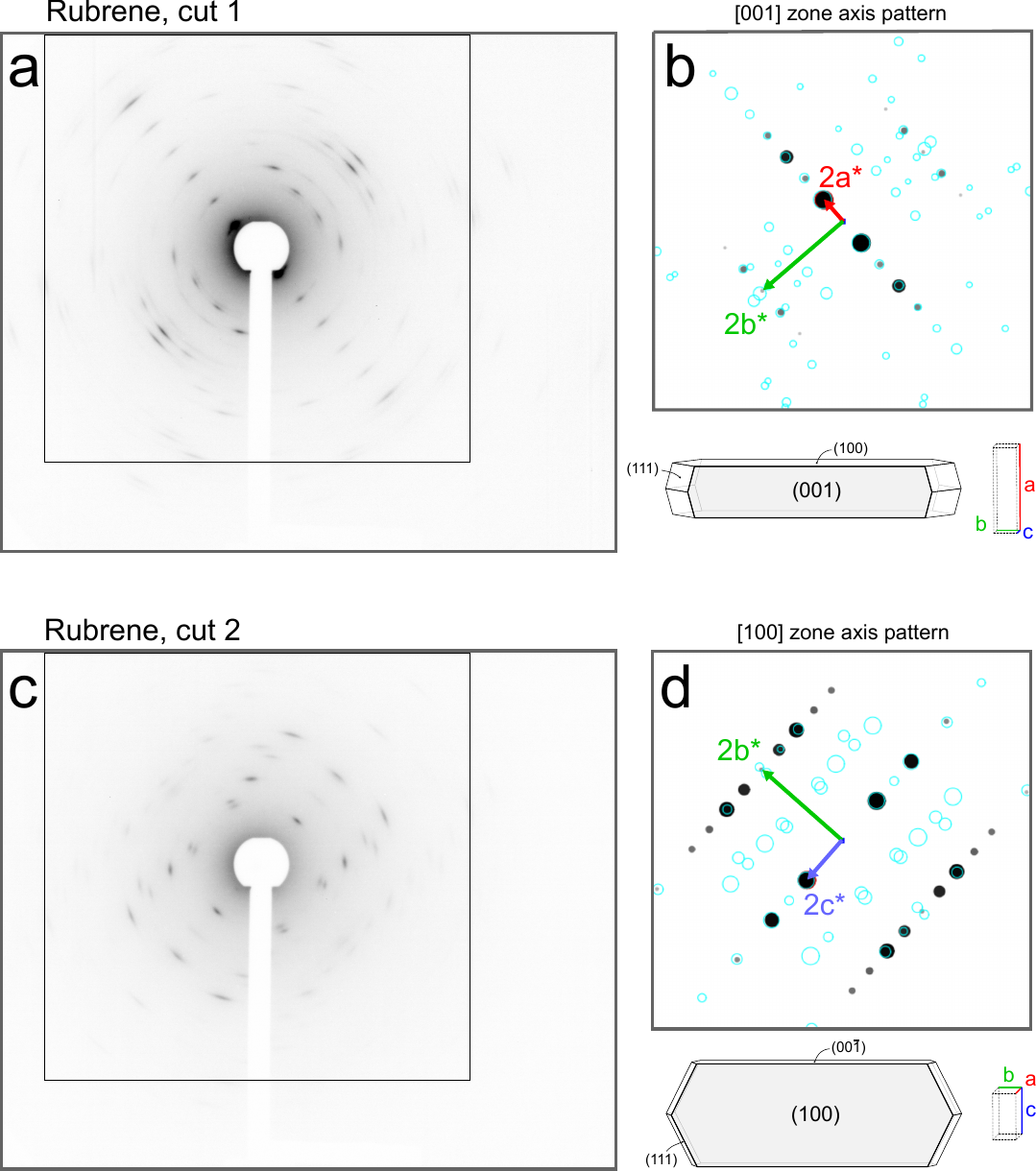}
\end{center}

\textbf{Diffraction images of two rubrene slices and their presumed crystal orientation}\\
\textbf{(a, c)} Diffraction images in linear intensity scaling as recorded with a 2560 x 2160 pixel detector (outer frame); the quadratic data region used by \textsc{Garfield} is indicated and identical to the region in panels b, d. \textbf{(b, d)} Zone-axis patterns for the presumed crystal cuts generated with \textsc{Garfield} using a small orientation spread of 0.1°; below the patterns are schematic drawings of orthorhombic rubrene crystals with growth facets according to Moret and Gavezzotti (2022) \cite{Moret2022}, illustrating the orientation of the slices relative to the unit cell of the crystal.

\newpage

\subsubsection{Figure Smp3B: \quad Results of \textsc{GridScan}}
\vspace{0.5in}

\begin{center}
\includegraphics[width=5.0in]{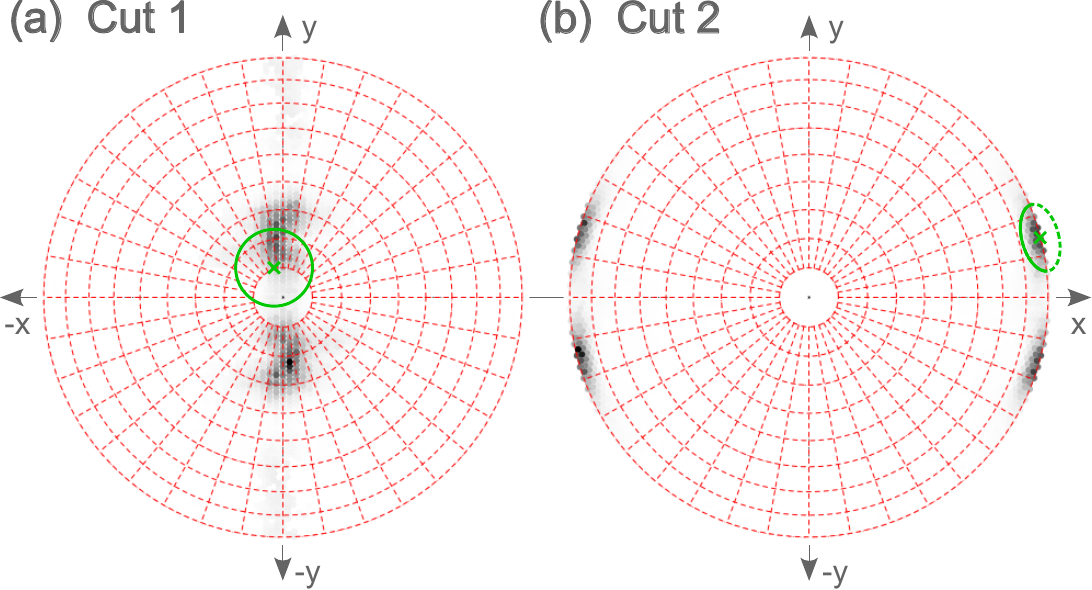}
\end{center}

\textbf{Orientations of rubrene slices suggested by \textsc{GridScan}}\\
Polar plots of electron beam directions relative to the unit cell, weighted by \textsc{GridScan}'s figure of merit (FOM) measuring the best match achieved with each beam direction. Only the upper hemisphere ($z \geq 0$) is shown; due to the symmetry of the crystal, the map looks virtually the same on the $z \leq 0$ hemisphere. $x,y,z$ are the directions of the standard Cartesian coordinate system associated with the unit cell of the crystal: $x || a,~ z || c^*$.
The dark regions indicate significant deviations of the crystal slices from the exact zone-axis orientation, which correspond to the center point in \textbf{(a)}, and to point $\Theta = 90^{\circ}, \Phi = 0$ on the $x$ axis in \textbf{(b)}. 

The beam directions corresponding to the best solutions found with  \textsc{GeoFit} (see Figures Smp3C and Smp3D) are marked with green crosses. The green ellipses around the crosses roughly delineate the range of orientational spread as determined by fitting the apparent mosaicity with ANISO modeling: fitted $\sigma_i^\mathrm{mos}$ values were 12.7° and 13.3° for cut 1, and 7.4° and 9.0° for cut 2. The third sigma value, describing the rotational disorder around the incident electron beam, was not fitted, but fixed and set to 3° for cut 1 and 2.5° for cut 2. Note that the center point in \textbf{(a)} is in the one sigma range, showing that a considerable part of slice 1 is not far from exact zone-axis orientation.

\newpage

\subsubsection{Figure Smp3C: \quad Slice 1, observed and simulated diffraction}
\vspace{0.025in}

\begin{center}
\includegraphics[width=5.0in]{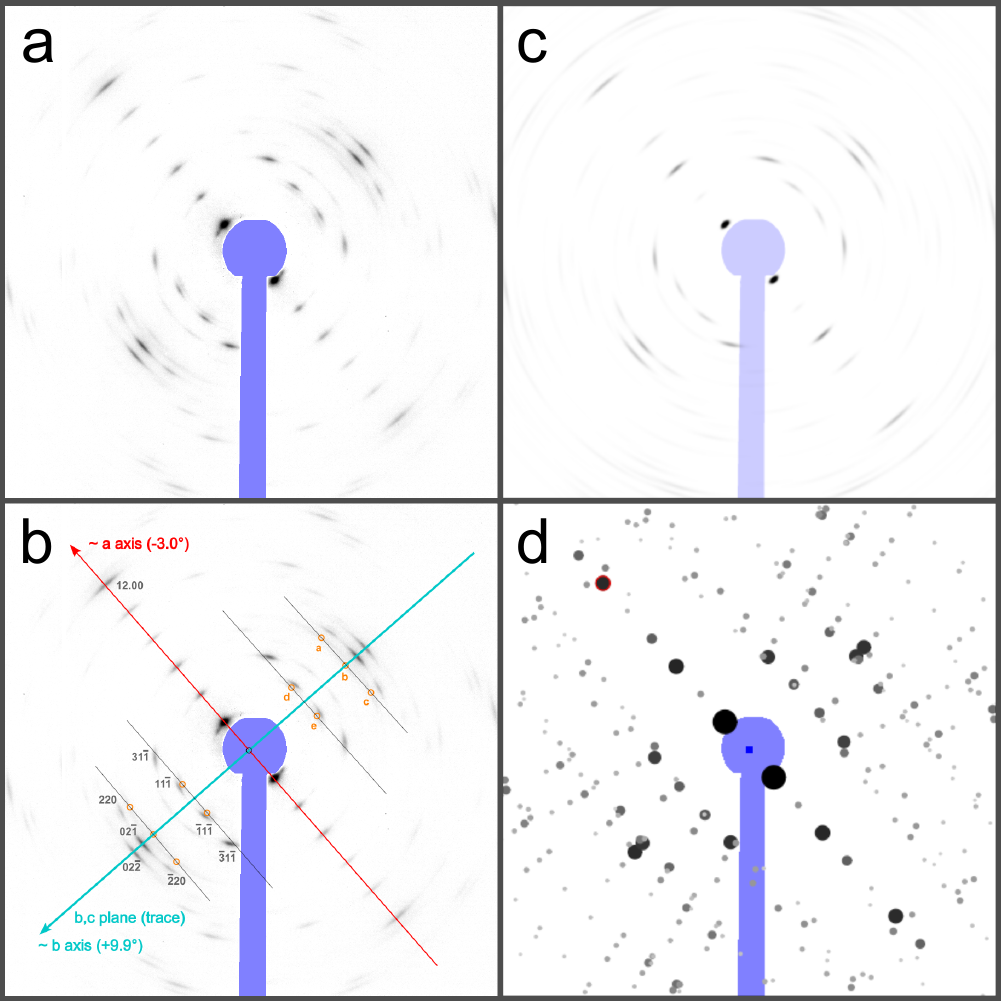}
\end{center}

\textbf{Observed diffraction and simulation with \textsc{Garfield}: ~Slice 1}\\
\textbf{(a)} Observed diffraction as in Figure Smp3A, a (square region) after background subtraction and with contrast enhancement (high-intensity cutoff at 60\% and gamma correction with $\gamma$ = 3.31). \textbf{(b)} same as (a) with annotations. \textbf{(c)} Simulated diffraction \textit{image} with the same contrast enhancement as in (a). \textbf{(d)} Simulated diffraction \textit{pattern}, i.e. schematic representation of the diffraction provided by \textsc{Garfield} to help identify individual reflections.

The red and cyan arrows in \textbf{(b)} are projections of axes $a$ and $b$ (or $a^*$ and $b^*$). Their deviations from the drawing plane are given in parentheses. A "+" sign ("-" sign) means that the axis points into the half-space above (below) the drawing plane. Bragg reflections (hkl) can be assigned unambiguously (up to symmetry equivalents) to almost all observed diffraction spots. However, a clear explanation is still missing for weak spots (Friedel pairs) denoted with letters a – e (see discussion below).

\newpage

\subsubsection{Figure Smp3D: \quad Slice 2, observed and simulated diffraction}
\vspace{0.025in}

\begin{center}
\includegraphics[width=5.0in]{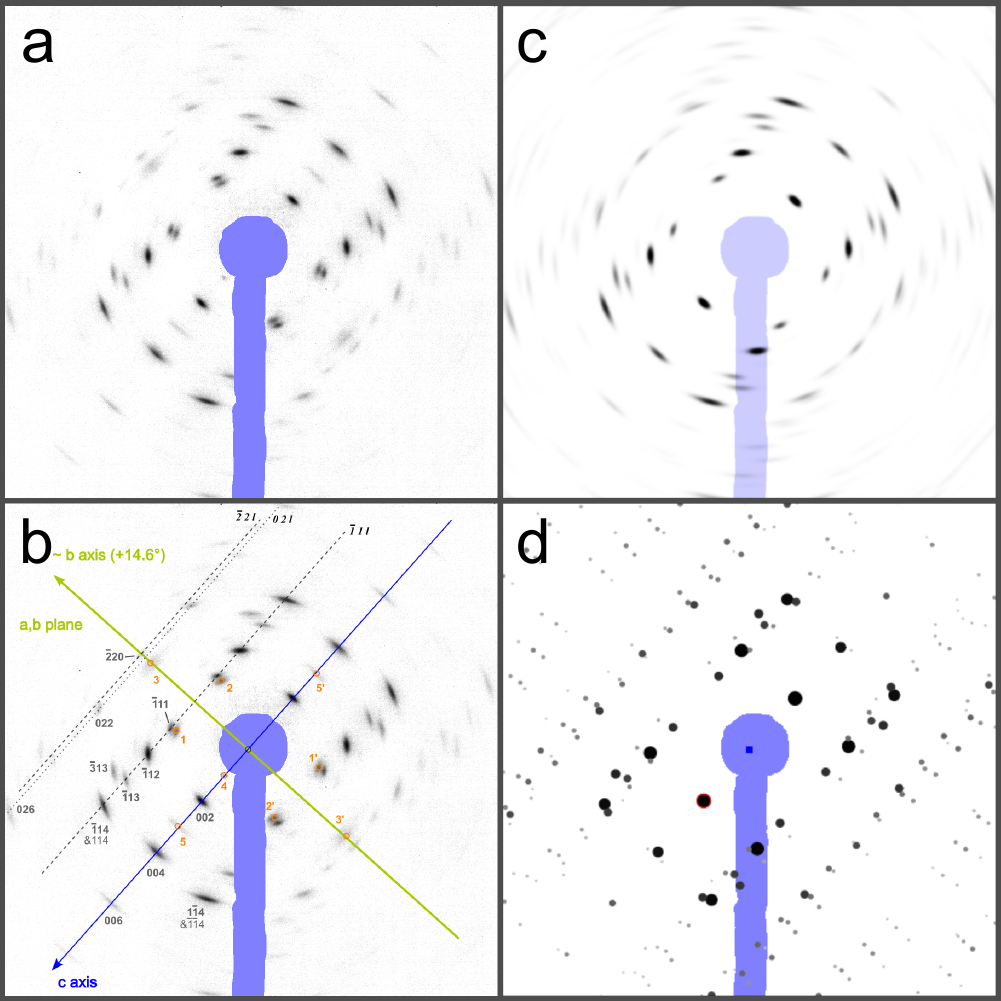}
\end{center}

\textbf{Observed diffraction and simulation with \textsc{Garfield}: ~Slice 2}\\
\textbf{(a)} Observed diffraction of slice 2, after background subtraction and with contrast enhancement as in Figure Smp3C, a. \textbf{(b, c, and d)}: analogous to Figure Smp3C, b, c, d.

As with diffraction from slice 1 (Figure Smp3C), most of the observed reflections can be indexed without ambiguity (up to symmetry equivalent solutions), but again, a few reflections marked with orange circles and denoted 1 -- 5 (Friedel mates 1', 2', 3', and 5'; 4' is masked)  are lacking a straightforward explanation.

\newpage

\subsubsection{Discussion: Unexpected spots}

As shown in Figures Smp3C and Smp3D, the simulation of electron diffraction with \textsc{Garfield} gives very satisfactory results, even if the absolute intensities are not reproduced very accurately: the "conventional R factors" calculated with 63 and 37 observed "Bragg spots", respectively, (i.e. partial intensities of Bragg reflections that possibly overlap) are 26\% for slice 1 and 16\% for slice 2. Apart from this, most of the fine details which only appear at high contrast in the observed diffraction and which were not included in the input for \textsc{Garfield} have a counterpart in the simulation. Thus, there is little doubt that the assumption of an orthorhombic Cmce crystal structure and the orientation and indexing found with \textsc{Garfield} are correct. 

However, a detailed comparison of simulated and observed diffraction also revealed the presence of unexpected reflections, suggesting that some features of the rubrene slices have not yet been properly captured. The unexpected reflections appear in both slices at positions compatible with the symmetry of the main pattern (two perpendicular mirror lines), so it seems unlikely that the unaccounted for reflections are completely unrelated to the crystal structure of the slices.

In kinematical diffraction theory, reflection conditions for space group Cmce lead to systematic absence of reflections. Considering the large size of the unit cell and the fact that coherently scattering domains are relatively small (presumably much smaller than 50 nm, the nominal thickness of the slices) it seems conceivable that reflection conditions are violated due to the limited size of the domains (truncated unit cells). This assumption could explain reflections 4, 5 and 5' in Figure Smp3D, b. (Reflection 4' is not observed due to the beam stop.) Furthermore, double diffraction can contribute to the intensities of these forbidden reflections. However, other unexpected reflections in both slices cannot be easily explained in the same way, because there are no forbidden reflections that could appear at the observed positions. Some of the unexpected reflections could perhaps be explained by the presence of Cmce domains with completely different orientations (see Table S-I, orthorhombic), but this is also difficult to believe since the d-spacings of all possible reflections of the rubrene Cmce structure do not agree very well with the observed values.

In a study of rubrene polymorphism by lattice phonon Raman microscopy \cite{Socci2017}, the coexistence of orthorhombic and triclinic polymorphs in optically homogeneous rubrene crystals was reported. Interestingly, the d-spacings of the triclinic form of rubrene (space group P$\overline{1}$, \cite{Huang2010, Matsukawa2010}) seem to agree much better with the d-spacings of the unexpected reflections (Table S-I, triclinic). 
Thus, phase mixing of orthorhombic and triclinic polymorphs might be a better explanation. If the relative orientations of the two phases are fixed by geometrical restraints of cocrystallization (in analogy to twinning laws), the fact that the diffraction patterns of unexpected and main reflections have the same symmetry would be a consequence of the growth form of orthorhombic rubrene "host" crystals, which reflects the mmm point symmetry of Cmce crystals. 

\newpage
\subsubsection{Table S-I: \quad d-spacings of unexpected spots}

\begin{table}[hb!]
    \footnotesize
    \centering
    \begin{tabular}{c c c c r r c r r r} 
    \noalign{\smallskip}
    \hline
    \noalign{\smallskip}
         \multicolumn{3}{c}{} &
         \multicolumn{3}{c}{triclinic (173 K)} &
         &
         \multicolumn{3}{c}{orthorhombic (293 K)} 
    \\
    \cline{4-6}
    \cline{8-10}
    \noalign{\smallskip}
         Slice & 
         Spots & 
         $d_\mathrm{obs.}$ &  
         \{hkl\} &  
         $d_\mathrm{hkl}$ &  
         $F_\mathrm{hkl}$ &
         &
         \{hkl\} &
         $d_\mathrm{hkl}$ &
         $F_\mathrm{hkl}$ 
    \\[12pt]
         \multirow{2}{*}{1}  & 
         \multirow{2}{*}{a, a', c, c'} & 
         \multirow{2}{*}{$3.71 \pm 0.08$} &  
         $1 \, \overline{2} \, \overline{1}$ &  
         3.701 &  
         4.760 &
         &
         $3 \, 1 \, 3$ &
         3.651 &
         113.414 
    \\
         & & &  
         $1 \, \overline{1} \, 2$ &  
         3.703 &  
         30.964 &
         &
         $5 \, 1 \, 2$ &
         3.697 &
         28.382 
    \\[12pt]
         \multirow{3}{*}{1}  & 
         \multirow{3}{*}{b, b'} & 
         \multirow{3}{*}{$3.88 \pm 0.09$} &  
         $1 \, 0 \, \overline{3}$ &  
         3.804 &  
         7.066 &
         & & &          
    \\
         & & &  
         $\mathbf{1 \, \overline{2} \, 0}$ &  
         \textbf{3.805} &  
         \textbf{32.323} &
         & & &          
    \\
         & & &  
         $1 \, 0 \, 2 $ &  
         3.862 &  
         7.111 &
         &
         $1 \, 1 \, 3$ &
         3.955 &
         37.043 
    \\[12pt]
         \multirow{3}{*}{1}  & 
         \multirow{3}{*}{d, d', e, e'} & 
         \multirow{3}{*}{$6.58 \pm 0.23 $} &  
         $0 \, 1 \, 1$ &  
         6.544 &  
         3.528 &
         &
         $2 \, 0 \, 2$ &
         6.357 &
         18.400 
    \\
         & & &  
         $1 \, 0 \, \overline{1} $ &  
         6.644 &  
         16.497 &
         & & &          
    \\
         & & &  
         $1 \, 0 \, 0 $ &  
         6.706 &  
         4.027 &
         &
         $^\mathrm{a}\, 4 \, 0 \, 0$ &
         6.715 &
         10.928 
    \\[18pt]
         \multirow{3}{*}{2}  & 
         \multirow{3}{*}{3, 3'} & 
         \multirow{3}{*}{$3.81 \pm 0.07 $} &  
         $1 \, 0 \, \overline{3}$ &  
         3.804 &  
         7.066 &
         &
         $5 \, 1 \, 2$ &
         3.697 &
         28.382 
    \\
         & & &  
         $\mathbf{1 \, \overline{2} \, 0}$ &  
         \textbf{3.805} &  
         \textbf{32.323} &
         & & &          
    \\
         & & &  
         $1 \, 0 \, 2 $ &  
         3.862 &  
         7.111 &
         &
         $1 \, 1 \, 3$ &
         3.955 &
         37.043 
    \\[12pt]
         \multirow{3}{*}{2}  & 
         \multirow{3}{*}{1, 1', 2, 2'} & 
         \multirow{3}{*}{$6.79 \pm 0.17 $} &  
         $1 \, 0 \, \overline{1}$ &  
         6.644 &  
         16.497 &
         & & &
    \\
         & & &  
         $1 \, 0 \, 0 $ &  
         6.706 &  
         4.027 &
         &
         $^\mathrm{a}\, 4 \, 0 \, 0$ &
         6.715 &
         10.928 
    \\
         & & &  
         $\mathbf{0 \, 1 \, \overline{1}}$ &  
         \textbf{7.106} &  
         \textbf{18.113} &
         &
         $0 \, 0 \, 2$ &
         7.217 &
         79.430 
    \\
    \noalign{\smallskip}
    \hline
    \noalign{\smallskip}
    \end{tabular}
    \caption*{

    \footnotesize

    \begin{tabular}{l p{5in}}
        \noalign{\smallskip}
        \multicolumn{2}{p{6in}}{\textbf{Unexpected diffraction spots and Bragg reflections with similar d-spacings }}
        \\
    \noalign{\smallskip}
        Spots & 
        {lists of unexpected reflections in rubrene slices 1 and 2 that are presumably crystallographically equivalent; spot labels as in Figures S3C and S3D; primed symbols denote Friedel mates. Note that spots labeled 4, 5, and 5' in Figure S3D are not listed as these can be explained by other effects (double diffraction or truncation of unit cells).}
        \\
    \noalign{\smallskip}
        $d_\mathrm{obs.}$ & 
        {measured mean d-spacing (in \r{A}) of the reflections listed in column Spots. Error margins were estimated by assuming $\pm3$ pixels uncertainty in the distance from the central beam and 1\% error in spatial calibration, using calibration factor and distortion fitted with \textsc{GeoFit}.}
        \\
    \noalign{\smallskip}
        \{hkl\} &
        {sets of equivalent lattice planes with d-spacings $d_\mathrm{hkl}$ close to  $d_\mathrm{obs.}$, calculated with crystal structure parameters for triclinic and orthorhombic rubrene phases. Few lattice planes with very small structure factors, $F_\mathrm{hkl} < 1 \, $\r{A}, have been excluded. } 
        \\
    \noalign{\smallskip}
        $d_\mathrm{hkl}$ &
        {d-spacing of \{hkl\} lattice planes in \r{A}} 
        \\
    \noalign{\smallskip}
        $F_\mathrm{hkl}$ &
        {structure factors for electrons in \r{A}, $F_\mathrm{000}$ = 120.18~\r{A} (triclinic), 480.70~\r{A}} (orthorhombic)
        \\
    \noalign{\smallskip}
        triclinic &
        {structure of triclinic rubrene measured at 173 K \cite{Huang2010}, Cambridge Structural Database CCDC code 726176; values calculated for the triclinic form as determined by Matsukawa et al. (2010) \cite{Matsukawa2010} are almost identical.}
        \\
    \noalign{\smallskip}
        orthorhombic &
        {293 K structure of orthorhombic rubrene by Jurchescu et al. (2006) \cite{Jurchescu2006}}
        \\
    \noalign{\smallskip}
        \textbf{bold face} &
        {lattice planes of the triclinic form that could \textit{consistently} (taking the relative orientation of the crystal cuts into account) explain all three Friedel pairs listed for slice 2 and one of the Friedel pairs listed for slice 1 are highlighted in bold face.}
        \\
    \noalign{\smallskip}
    \noalign{\smallskip}
        $^\mathrm{a}$ &
        {Reflections of the \{400\} lattice planes can be excluded as mutch stronger \{200\} reflections should also be observed, which is not the case.}
    \end{tabular}

    }

    \label{tab:rubrene}
\end{table}

\printbibliography[
heading=bibintoc,
title={References}
]